\documentclass[12pt]{article}
\usepackage{color,amssymb,epsfig,verbatim}
\usepackage{graphicx} 
\usepackage{bm}
\usepackage{multirow}
\usepackage{amsmath}
\usepackage{lscape}
\usepackage[table,xcdraw]{xcolor}
\usepackage{natbib}
\usepackage{hyperref}
\usepackage{algorithm}
\usepackage{algorithmic}
\usepackage{accents}
\newtheorem{proposition*}{Proposition}
\usepackage[utf8]{inputenc}
\usepackage{subcaption}
\usepackage{diffcoeff}
\usepackage{derivative}

\usepackage[figuresright]{rotating}
\hypersetup{colorlinks,linkcolor={black},citecolor={gray},urlcolor={red}}

\def\log{\hbox{log}}

\def\boxit#1{\vbox{\hrule\hbox{\vrule\kern6pt
          \vbox{\kern6pt#1\kern6pt}\kern6pt\vrule}\hrule}}

\def\bse{\begin{eqnarray*}}
\def\ese{\end{eqnarray*}}
\def\be{\begin{eqnarray}}
\def\ee{\end{eqnarray}}
\def\bq{\begin{equation}}
\def\eq{\end{equation}}
\def\bse{\begin{eqnarray*}}
\def\ese{\end{eqnarray*}}

\def\w{{\bf w}}

\usepackage[para]{threeparttable}
\usepackage{rotating}
\usepackage[mathlines, displaymath]{lineno}
\linenumbers

\begin{document}
\nolinenumbers
\thispagestyle{empty}

\hfill\today \\ \\

\baselineskip=28pt
\begin{center}
{\LARGE{\bf High-dimensional Bayesian Model for Disease-Specific Gene Detection in Spatial Transcriptomics }}
\end{center}
\baselineskip=14pt
\vskip 2mm
\begin{center}
Qicheng Zhao and Qihuang Zhang\footnote{\baselineskip=10pt Corresponding Author: Department of Epidemiology, Biostatistics and Occupational Health, McGill University, Montreal, Quebec, Canada H3A 1G1, qihuang.zhang@mcgill.ca}

\end{center}
\bigskip

\vspace{8mm}

\begin{center}
{\Large{\bf Abstract}}
\end{center}
\baselineskip=17pt
{ Identifying disease-indicative genes is critical for deciphering disease mechanisms and has attracted significant interest in biomedical research. Spatial transcriptomics offers unprecedented insights for the detection of disease-specific genes by enabling within-tissue contrasts. However, this new technology poses challenges for conventional statistical models developed for RNA-sequencing, as these models often neglect the spatial organization of tissue spots. In this article, we propose a Bayesian shrinkage model to characterize the relationship between high-dimensional gene expressions and the disease status of each tissue spot, incorporating spatial correlation among these spots through autoregressive terms. Our model adopts a hierarchical structure to facilitate the analysis of multiple correlated samples and is further extended to accommodate the missing data within tissues. To ensure the model's applicability to datasets of varying sizes, we carry out two computational frameworks for Bayesian parameter estimation, tailored to both small and large sample scenarios. Simulation studies are conducted to evaluate the performance of the proposed model. The proposed model is applied to analyze the data arising from a HER2-positive breast cancer study.

}

\vspace{8mm}

\par\vfill\noindent

\par\medskip\noindent
\underline{\bf Short title}: Bayesian Model for Disease-Specific Gene Detection

\clearpage\pagebreak\newpage
\pagenumbering{arabic}

\newlength{\gnat}
\setlength{\gnat}{22pt}
\baselineskip=\gnat

\clearpage



\section{Introduction} 
Complex diseases, such as cancer \citep{warburg1956origin}, neurodegenerative disorders \citep{martin1999molecular}, and autoimmune diseases \citep{dominguez2018regulatory}, often involve dysregulation at the cellular level that may not be uniformly distributed across tissues. To decipher the functions of cells in multicellular organisms and the mechanisms of complex disease, understanding the relationship between high-dimensional gene expression data and disease status is crucial. The initial sequencing approaches were conducted on bulk tissue \citep{li2021bulk}. Although economical, these approaches lacked the resolution to capture the heterogeneity of disease status within tissues. In contrast, single-cell RNA-seq (scRNA-seq) allows for the profiling of gene expression in a single-cell resolution, providing a within-tissue contrast of disease status \citep{sant2023approaches}. Such a contrast naturally introduces a matched design, bringing the opportunity to control for potential confounders. Recently, the advances in spatial transcriptomics \citep{staahl2016visualization} further enable maintaining the spatial context of the tissue while profiling gene expressions. Its technical procedure involves capturing and sequencing RNA \textit{in situ}, thereby preserving the spatial organization of the cellular gene expressions within the tissue architecture.

With the emergence of these novel types of data, identifying informative genes that can decipher biological mechanisms has become a keen area of interest. One key category of these genes is \textit{spatially variable genes}. Within this scope, the current works \citep[etc.]{hao2021somde, svensson2018spatialde, jiang2023sinfonia} highlight the importance of genes in understanding spatial heterogeneity within tissues and their potential in revealing complex biological insights that traditional transcriptomics techniques cannot provide. Such a framework identifies genes that capture spatial variation information and is therefore used as a crucial preprocessing step to reduce the number of genes before performing various downstream tasks involving spatial variation, such as tissue segmentation \citep{li2022bass}, cell recovery \citep{zhang2023leveraging}, and resolution enhancement \citep{zhang2024inferring}. On the other hand, while these detected genes are strongly representative across spatial domains, they are not necessarily indicative of diseases. Our focus is on the second type of interesting genes, \textit{disease-specific genes}. They are crucial in understanding the relationship between high-dimensional gene expression data and disease status, and thus informing insights for diagnosis and therapeutic strategies. Identifying disease-specific genes has been extensively studied using bulk RNA-seq data, including works by \cite{bauer2008walking}, \cite{jia2011dmgwas}, \cite{guala2014maxlink}, \cite{leiserson2015pan}, and scRNA-seq data \citep{ruan2021disnep}, but it is scarcely explored in spatial transcriptomics. Motivated by this, we propose a high-dimensional Bayesian model to efficiently detect the disease-specific genes leveraging the spatial structure of transcriptomics data. This model can be adapted for new tissue samples, serving as a predictive tool to identify disease regions and assist in diagnosis. 

 There are two primary methods of generating spatial transcriptomics data: \textit{sequencing-based} and \textit{imaging-based}. Sequencing-based spatial transcriptomics profiles gene expression counts via next-generation sequencing (NGS) technology. For example, 10X Visium is a commercialized platform designed to generate spatial transcriptomics data, capturing gene expression profiles across various spots regularly spaced around the collected tissues, while preserving their spatial locations. Despite the relatively low resolution of this data, as each of the spots usually contains multiple cells, it remains a cost-effective solution for spatially resolved transcriptomics and often provides histological images. In contrast, image-based spatial transcriptomics, such as multiplexed error-robust fluorescence \textit{in situ} hybridization (MERFISH), offers higher resolution and can capture cellular or even subcellular organization \citep{williams2022introduction}. Since disease organization information is often provided through tissue annotations based on histological images, in this article, we mainly focus on sequencing-based spatial transcriptomics, given that histological information is typically lacking in imaging-based methods. Besides, the affordability of the sequencing-based spatial transcriptomics method makes it a practical choice for large-scale, multi-site and cross-tissue studies.

Our proposed model addresses three challenges imposed by the special data structure of spatial transcriptomics: high dimensionality, missing data, and computation burden.

Firstly, disease mechanisms can be driven by the simultaneous interaction of multiple genes, this requires us to consider modeling the additive effects of these genes collectively. That is, our focus will be on the conditional effects of each gene in the presence of others, which is extremely challenging due to the high dimensionality of the data. At present, dimension reduction techniques have been introduced for analyzing gene expressions \citep[e.g.,][]{feng2020dimension, sorzano2014survey}. However, the existing methods for reducing dimensions mainly focus on a specific goal such as extracting features for clustering, which are not general enough for identifying disease-specific genes, let alone for handling data with spatial correlations. To fill in this gap, we employed a Bayesian shrinkage model to effectively manage the association analysis in high-dimensional data.

Secondly, the presence of missing data makes the inference of autologistic model to be more challenging as the disease status are appearing as both predictors and responses, leading to the complete case analysis to be impossible. Moreover, the problem of the missing spots are entirely missing, meaning that neither disease outcome nor gene expression is captured on these spots, which has not been extensively studied in spatial transcriptomics \citep{lopez2019joint}. Unlike the missingness in scRNA-seq, missing spots often result from sample collection procedures or experimental errors, which may cause the missingness subject to spatial correlations, each leading to a more complicated missing data mechanism. In this article, we discuss missing data mechanisms to be ignorable or nonignorable, and address scenarios where the missing mechanism involves autocorrelation among adjacent spots.

Thirdly, spatial transcriptomics data often include a large number of spots, genes, and slices. With the decreasing costs of generating new datasets, it is crucial to ensure that the methods are scalable to accommodate large datasets. However, the widely used Bayesian sampling methods, including the No-U-Turn Sampler (NUTS) by \cite{r2}, have been demonstrated to be excessively computational costly, particularly for high-dimensional data. While these sampling techniques yield the exact posterior distributions and robust uncertainty quantification, their computational demands hinder them for large-scale analyses. Alternatively, variational inference efficiently approximates posterior distributions by optimizing a set of parameters. In this work, we develop both the NUTS and a variational inference approach. These dual options allow users to trade off based on their priority between precision and computational efficiency during implementation.

The remainder of the article is organized as follows. In Section \ref{sec:model}, we present the notation and model setup for the scenarios involving single slice, multiple slices, and missing data, respectively. Section \ref{sec:Methodology} outlines the Bayesian inference framework for each model. Then, we extend the strategies for handling missing data in Section \ref{ModelMiss}. Section \ref{sec:estimation} details the computational approach for the parameter estimation. In Section \ref{sec:Simulation} we validate the performance of our models through simulation studies.  Section \ref{sec:Application} demonstrates the application of our model to data arising from HER2-positive breast cancer study \citep{r5}. Finally, we conclude with a discussion in Section \ref{sec:Discussion}.



\section{Model Setup}\label{sec:model} 

\subsection{Main Model}

For $i=1,\dots,n$, let $Y_i$ $\in \{0,1\}$ be a binary variable to denote the disease status captured on spot $i$ with a two-dimensional location coordinate $s_i$. For $i=1,\dots,n$, consider $d$-dimensional covariates  $X_i$ = $(X_{i1}, ..., X_{id})^{\intercal}$ with $X_{ik}$  denoting the expression level of gene $k$ on spot $i$. Let $\mu_i = \text{E}(Y_i)$. We consider the association of $Y_i$ and $X_i$ is postulated by the model
\begin{eqnarray} \label{eqn:general_equation}
\text{logit}(\mu_i)  = \beta^{\intercal} X_i+\sum_{j\neq i}\eta_{ij}Y_j,
\end{eqnarray}

\noindent where $\beta$ is the associated covariate parameter of primary interest, logit(c) = log($\frac{c}{1-c}$) and $\eta_{ij}$ are dependence parameters characterizing the correlations between the outcomes observed at different spots.

Without extra constraints, model identifiability issue arises as the number of parameters is greater than $n$. Hence, assumptions are often made to reduce the number of parameters. Here, we consider restricting the correlations with spots $i$ into its neighbor set, $\mathcal{N}(i)=\left\{j: ||s_i - s_j||_2 \leq \delta \right\}$, with $s_i$ and $s_j$ being the locations of spots $i$ and $j$, respectively, and $\delta$ being a pre-specified threshold. This gives a pair-wise dependencies assumption, 
\begin{equation*} \label{pair-wise}
    P(Y_i|\Tilde{Y_r},\Tilde{Y_t})=P(Y_i|\Tilde{Y_r}),
\end{equation*}
\noindent where $\Tilde{Y_r} = \left\{Y_r,r\in \mathcal{N}(i)\right\}$, $\Tilde{Y_t} = \left\{Y_t,t\notin \mathcal{N}(i)\right\}$ and the threshold $\delta$ is often taken as 1, i.e., the neighbor set is taken as the most adjacent spots. Further isotropic assumptions can be considered (e.g., \citealp{hughes2011autologistic}) for (\ref{eqn:general_equation}), where  $\eta_{ij}=\frac{\eta}{|\mathcal{N}(i)|}>0$, for $i, j=1,\dots, n$, representing the average autocorrelation factor contributed by one neighbor spot, and $|\mathcal{N}(i)|$ is the number of spots included in the neighbor set of spot $i$, to adjust for the boundary effect to be discussed in Section \ref{model:missing_data_model}.

Considering both pair-wise dependencies and isotropic assumptions, stemming from (\ref{eqn:general_equation}), a restricted autologistic model is given by
\begin{eqnarray} \label{restriautomodel}
\text{logit}(\mu_i)  = \beta^{\intercal} X_i+\frac{\eta}{|\mathcal{N}(i)|} \sum_{j \in \mathcal{N}(i)}Y_j.
\end{eqnarray}

We comment that in the field of spatial transcriptomics, both the pair-wise dependencies and isotropic assumptions are frequently considered. For example, \cite{zhu2018identification} considered the spatial dependencies only depend on immediate neighboring spots. On the other hand, the proposed model (\ref{eqn:general_equation}) has the room to accommodate dependencies at greater distances by modulating the value of $\delta$. The isotropic assumption stems from biological homogeneity, suggesting that in various tissues, the absence of a predominant direction of gene expression might be anticipated. Furthermore, with many sequencing-based gene expression profiling techniques such as 10X Visium, the resolution might not be sufficiently high enough to discern fine-grained directional patterns.

\subsection{Joint Modelling of Multiple Slices} \label{model:mult_slice_model}

In spatial transcriptomics, it is common that multiple slices from the same tumor tissue are collected and sequenced, which are often referred to as biological replicates. For instance, 10X Visium platform allows for generating spatial transcriptomic data for four serial slices (\citealp{marx2021method}), enabling statistical analysis with a larger sample size. However, in such a scenario, batch effects may be the main concern and thus need to be adjusted for via statistical modeling. 

In this section, we extend the model (\ref{restriautomodel}) to jointly model the multiple slices. We first introduce $c=1,\dots, C$ to denote the index of tissue donors, where $C$ is the total number of tissue donors. Then, for participant $c$, a total of $G_c$ slices are collected and we index each tissue slice by $g_c = 1,\dots, G_c$. Without the loss of generality, we consider $G_1=G_2=\dots=G_c=G$ and hence suppress the subscript $c$ in $g_c$ into $g$. We further incorporate these index into $X_i$ and $Y_i$ to indicate their source of slices, resulting in $X_{cgi}$ and $Y_{cgi}$, respectively. Consider the following model,
\begin{eqnarray} \label{eqn:general_equation_ms}
\text{logit}(\mu_{cgi})  = \beta^{\intercal} X_{cgi}+\frac{\eta_{cg}}{|\mathcal{N}(i)|}\sum_{j \in \mathcal{N}(i)} Y_{cgj} + U_{cg},
\end{eqnarray}

\noindent where $U_{cg}$ is a random effect variable. We assume $U_{cg}$ follows $N(0,\Sigma_{c})$ to account for correlation among slices within an individual, where the $(s,t)$th element of $\Sigma_c$ is $\rho_{st}\sigma_{c}^2$. Hence, the covariance matrix $\Sigma_c$ can be fully parameterized. Such a specification implicitly assumes $U_{c_{1}g_{1}}$  and $U_{c_{2}g_{2}}$ are independent for any $(g_1,g_2)$ of $c_1 \neq c_2$.

To account for the nature of correlation structure among slices from the same donor, we consider an exchangeable structure assuming that the pairwise correlations are the same, i.e., for $\forall s \neq t$, $\rho_{st}=\rho$ and for $\forall s=t$, $\rho_{st}=1$. In some instances, tissue slices are sequentially obtained from a larger specimen. Consequently, it is natural to consider the correlation between two adjacent slices influenced by their relative positions. Under such circumstances, an autoregressive correlation structure would be more appropriate, where $\rho_{st}=\rho^{|g_s - g_t|}$ for slices $s$ and $t$.

In practice, while multiple slices might be sourced from the same donor, their processing pipeline is often the same. This consistency allows for the assumption of consistent spatial effects, leading to the constraint that $\eta_{cg}=\eta$ for each $c$ and $g$.

\subsection{Boundary Effect and Missing Spots}\label{model:missing_data_model}

Model (\ref{eqn:general_equation}) assumes all $Y_j$ to be observed. However, $Y_j$ can be unknown under two scenarios. The first one occurs for a spot (e.g., Spot $i$) on the outer edges. In this case, its neighboring spot (e.g., Spot $j$) may be unobserved because of the natural boundary of the tissue. Consequently, the spots on the boundary will have fewer number of neighbors than others. Therefore, instead of just using a single parameter to account for autocorrelations $\eta_{ij}=\eta$ (e.g. \citealp{hughes2011autologistic}), we adjust it by the number of neighbors in (\ref{restriautomodel}). 

On the other hand, for the missing spots isolated in the interior of the tissue, their missingness often results from sample collection procedure. This missing cannot be ignored because the corresponding $\eta_{ij}$ might be non-zero. Without a proper adjustment, ignoring them will lead to the parameter estimation of (\ref{eqn:general_equation}) to be biased as it is equivalent to forcing $\eta_{ij}$ to be 0. In order to address this, we introduce new notations, $Y_{j}^{obs}$ and $Y_{j}^{mis}$, respectively, to distinguish the observed and unobserved disease status for spot $j$, and a binary variable $R_j$ to indicate the observation status of $Y_j$. Then, model (\ref{restriautomodel}) can be rewritten as
\begin{eqnarray} \label{ultimate}
\text{logit}(\mu_i)  = \beta^{\intercal} X_i+\frac{\eta}{\mathcal{N}(i)} \sum_{j \in \mathcal{N}(i)}\left\{Y_j^{obs} R_j+{Y_j}^{mis} (1-R_j)\right\},
\end{eqnarray}

\noindent where $R_j$ is 1 if $Y_j$ is observed, and otherwise, it would be 0. 

Because biological tissue is naturally continuous and interconnected, the failure of collecting one spot during the slice-collection procedure might increase the chance of the adjacent spot being missed as well, leading to the missing pattern tending to be autocorrelated. Hence, we consider an autocorrelated missing pattern:
\begin{eqnarray*} \label{model:assumption_autocorrelated_missing}
R_i \not\perp R_j \ \text{and} \ R_i \bot R_k|R_j \ \text{for } \ k\notin \mathcal{N}(i) \ \text{and} \ j \in \mathcal{N}(i).
\end{eqnarray*}

However, such a complex missing pattern might be \textit{ignorable} under certain circumstances. For example, the missing spots are often due to errors in the experimental operations such as tissue slicing, where the missingness is independent of the data observed on the tissue. Under this assumption, we consider the missing mechanism to be \textit{missing completely at random} (MCAR). That is to say, for $i = 1,\dots,n$, $P(R_i|\tilde{X},\tilde{Y})=P(R_i)$, where $\tilde{X} = (X_1^{\intercal}, \dots, X_n^{\intercal})^\intercal$ and $Y = (Y_1, \dots, Y_n)^{\intercal} $.

On the other hand, the experimenter's bias can influence the bias pattern, leading to the missing occurring differently according to the disease outcome. For example, the experimenter might have less incentive to include the normal regions and therefore process them with less care, which causes the control region to have a different missing rate compared to the disease group. In this scenario, the ignorable assumption may be no longer reasonable. Instead, we model the missing status of spot $i$ by its outcome as well as its neighboring missing status: 
\begin{eqnarray} \label{model:nonignorable_missing_model}
\text{logit} \ P(R_i=1)  = \gamma_0+\gamma_1 Y_i+\frac{\gamma_2}{|\mathcal{N}(i)|} \sum_{j\in \mathcal{N}(i)}R_j,
\end{eqnarray}

\noindent where $\gamma=(\gamma_0,\gamma_1,\gamma_2)^{\intercal}$ are associated parameters. The values of $\gamma_1$ and $\gamma_2$ determine the complexity of the missingness mechanism. When $\gamma_2 \neq$ 0, the missing mechanism is subject to spatial autocorrelation. However, when $\gamma_1=0,$ the missingness degenerates into an ignorable form, making model (\ref{model:nonignorable_missing_model}) unnecessary and $\gamma_2$ to become irrelevant.

\section{Methodology}\label{sec:Methodology}
\subsection{Spike and Slab Autologistic Model } \label{bayesian1}

To fulfill variable reduction, we integrate spike and slab priors with spike component specified as absolutely continuous spike into the following Bayesian hierarchical models:
\begin{equation}
\begin{aligned} \label{SSM}
(Y_i| X_i) \stackrel{i.i.d}{\sim} \text{Bernouli}(\mu_i),  i=1,\dots,n,
\\ \eta \sim \text{Uniform}(0, c_1),
\\(\beta_k|\iota_k , \tau_k^2) \stackrel{i.i.d}{\sim} \text{Normal}(0,\iota_k \tau_k^2), 
\\(\iota_k|v_0, w) \stackrel{i.i.d}{\sim} (1-w)\delta_{v_0}(\cdot) + w\delta_1(\cdot),
\\({\tau_k}^{-2}|b_1,b_2) \stackrel{i.i.d}{\sim} \text{Gamma}(b_1,b_2), k=1,\dots,d,
\\w \sim \text{Uniform}[0,1],
\end{aligned}
\end{equation}

\noindent where $\mu_i$ is modelled by (\ref{eqn:general_equation}), $c_1$, $v_0$, $b_1$, $b_2$ are hyperparameters to be prespecified, $\delta_v(\cdot)$ function is a discrete measure concentrated at value $v$, and the choice for $v_0$ is usually a small value near zero. We comment that this framework is also called \textit{Normal Mixture of Inverse Gamma } (NMIG) proposed by \cite{r1}, which takes full advantage of the conjugate Inverse Gamma prior for $\tau_k^{2}$ and leads to highly efficient computation when applied to high-dimensional datasets.

\subsection{Model with Multiple Tissue Slices Available}\label{model_for_multiple_slice}

To accommodate the scenarios where datasets are collected with multiple replicates of slices from different individuals, as considered in Section \ref{model:mult_slice_model}, we extend the model (\ref{SSM}). The data-generating process for $Y_i|X_i$ is now indexed by $c,g$ and $i$: 
\begin{equation*}
Y_{cgi}| X_{cgi} \stackrel{i.i.d}{\sim} \text{Bernouli}(\mu_{cgi}),
\end{equation*}

 \noindent for $c=1,...,C, g=1,\dots,G, i=1,\dots,n$, where $\mu_{cgi}$ is modeled by (\ref{eqn:general_equation_ms}). Model (\ref{eqn:general_equation_ms}) introduces a latent random effect $U_{cg}$. Here, we further extend (\ref{SSM}) into a Bayesian hierarchical model by including 
\begin{equation*} \label{methodology:bayesian_framework_multip_slice_ssm}
U_{cg} \sim \text{MVN}(0,\Sigma_c),
\end{equation*}

\noindent where $\sigma^2$ in $\Sigma_c$ are assigned with priors 
\begin{equation*}
\begin{aligned}
& \sigma^{-2}_c \stackrel{i.i.d}{\sim} \text{Gamma}(b_3,b_4), \ \text{for}\ c=1,...,C, \ \text{and}
\\& \rho \sim \text{Uniform} (b_5, b_6).    
\end{aligned}
\end{equation*}

 \noindent Here, the parameters denoted as $b_3$, $b_4$, $b_5$, $b_6$ are characterized as hyperparameters to be specified.

\subsection{Bayesian Inference}\label{subs:bays.inf}
Inference about the parameter $\theta = (\eta,\beta,\iota,\tau,w)^{\intercal}$ for Section \ref{bayesian1} or  $\theta = (\eta_{cg},\beta,\iota,\tau,w,U_{cg},\sigma_c,\rho)^{\intercal}$ for Section  \ref{model_for_multiple_slice} are carried out by the following posterior distribution:
\begin{equation}
\begin{aligned}\label{Inference equation about beta}
f(\theta|y,x)=\frac{f(\theta,y|x)}{f(y|x)}\propto f(y|x,\theta)\pi(\theta),
\end{aligned}    
\end{equation}

\noindent where $f(\theta,y|x)$ is the joint distribution of $Y$ and $\theta$ conditional on $X$, $\pi$($\theta$) is the prior distribution of parameter $\theta$. 

For scenarios described in Section \ref{bayesian1} or \ref{model_for_multiple_slice}, $f(y|x,\beta)$ is given by (\ref{SSM}), and $f(y|x)$ = $\int f(y|x,\theta)\pi(\theta) d \theta$, with 
\begin{equation*}
\begin{aligned}\label{posterior}
\pi(\theta)&=(1-w) \times \frac{1}{\sqrt{2\pi v_0\tau^2}} \exp\left(-\frac{\beta^2}{2 v_0\tau^2}\right) \times \frac{b_2^{b_1}}{\Gamma(b_1)} {\tau}^{2(-b_1 - 1)} e^{-\frac{b_2}{\tau}} \times \frac{1}{c_1}+w \times \frac{1}{\sqrt{2\pi\tau^2}}\\
&\exp\left(-\frac{\beta^2}{2\tau^2}\right) \times \frac{b_2^{b_1}}{\Gamma(b_1)} {\tau}^{2(-b_1 - 1)} e^{-\frac{b_2}{\tau}} \times \frac{1}{c_1},
\end{aligned}    
\end{equation*}

\noindent  for Section \ref{bayesian1} and 
\begin{equation*}
\begin{aligned}
\pi(\theta)&=(1-w) \times \frac{1}{\sqrt{2\pi v_0\tau^2}} \exp\left(-\frac{\beta^2}{2 v_0\tau^2}\right) \times \frac{b_2^{b_1}}{\Gamma(b_1)} {\tau}^{2(-b_1 - 1)} e^{-\frac{b_2}{x}} \times \frac{1}{c_1} \times \frac{1}{(2\pi)^{G_c/2} |\boldsymbol{\Sigma}|^{1/2}}\\
&\exp\left( -\frac{1}{2} U_{cg}^\intercal \boldsymbol{\Sigma}^{-1} U_{cg}\right) \times \frac{b_4^{b_3}}{\Gamma(b_3)} {\sigma_c}^{2(-b_3 - 1)} e^{-\frac{b_4}{\sigma_c}} \times \frac{1}{b_6-b_5} +w \times \frac{1}{\sqrt{2\pi\tau^2}} \exp\left(-\frac{\beta^2}{2\tau^2}\right) \times\\
&\frac{b_2^{b_1}}{\Gamma(b_1)} {\tau}^{2(-b_1 - 1)} e^{-\frac{b_2}{\tau}} \times \frac{1}{c_1} \times \frac{1}{(2\pi)^{G_c/2} |\boldsymbol{\Sigma}|^{1/2}} \exp\left( -\frac{1}{2} U_{cg}^\intercal \boldsymbol{\Sigma}^{-1} U_{cg}\right) \times \frac{b_4^{b_3}}{\Gamma(b_3)} {\sigma_c}^{2(-b_3 - 1)} e^{-\frac{b_4}{\sigma_c}} \times \frac{1}{b_6-b_5},
\end{aligned}    
\end{equation*}
 for Section \ref{model_for_multiple_slice}.

 The posterior mean $\hat{\theta}=\text{E}(\theta|Y,X)$ is established as the Bayesian estimator for $\theta$. For the implementation of parameter estimation, a sequence of parameters will be firstly sampled from the posterior distribution given by (\ref{Inference equation about beta}), and then the sample mean of the resulting sequence will be computed as the Bayes point estimator $\hat{\theta}$ for $\theta$.

\section{Inference with Missing Data }\label{ModelMiss}
With the presence of missing data, the inference of the $\theta$ in Section \ref{subs:bays.inf} starts with the joint distribution of $(Y,R)$, given by
\begin{equation}\label{eqn:inf_miss}
    f(y,r|x,\theta,\gamma)=f(y|x,\theta)f(r|y,\gamma).
\end{equation}
Then, the inference based on the observed data is given by 
\begin{equation}\label{eqn:inf_obs_data}
    f(y^{obs},r|x,\theta,\gamma)=\sum_{y^{mis}}f(y,r|x,\theta,\gamma).
\end{equation}
When the data is \textit{missing completely at random} (MCAR), which means the missing data mechanism $R$ is completely independent of $Y$, (\ref{eqn:inf_miss}) can be further factorized as
\begin{equation*}
    f(y,r|x,\theta,\gamma)=f(y|x,\theta)f(r|\gamma).
\end{equation*}
This indicates that the likelihood $f(y|x,\theta)$ is not influenced by the likelihood of $f(r|\gamma)$, and hence we can ignore the model of $R$ when conducting the inference of $\theta$ and carry out the inference only via $f(y|x,\theta)$. Here, (\ref{eqn:inf_obs_data}) would become 
\begin{equation}\label{eqn:inf_obs_data_mcar}
    f(y^{obs}|x,\theta,\gamma)=\sum_{y^{mis}}f(y|x,\theta,\gamma).
\end{equation}
The inference on (\ref{eqn:inf_obs_data_mcar}) involves a summation over all potential binary configurations of $y^{mis}$, which can be computationally expensive when the number of missing spots is substantial. Therefore, we incorporate a data augmentation procedure by treating $y^{mis}$ as a latent variable and imputing them via iterative multiple imputation with a Gibbs sampler. This is equivalent to adjusting (\ref{SSM}) by considering the separated data generating process of outcome $Y_i$ according to whether spot $i$ is observed. Specifically, let $\mathcal{M}$ denote the index set of spots are missing,
\begin{equation}
\begin{aligned} \label{ultimate_fram}
Y_i^{obs}| X_i, Y_j^{obs}, {Y_j}^{mis}, R_j \stackrel{i.i.d}{\sim} \text{Bernouli}(\mu_i),  \ \text{for} \ i=1,\dots,n, j \in N(i),
\\{Y_k}^{mis}|Y_w^{obs},{Y_w}^{mis},R_w \sim \text{Bernouli}\left(\frac{1}{{|\mathcal{N}{(k)}|}}\sum_{w \in \mathcal{N}(k)}\left\{Y_w^{obs} R_w+{Y_w}^{mis} (1-R_w)\right\}\right), \ \text{for} \ k \in \mathcal{M},
\end{aligned}
\end{equation}

\noindent where $\mu_i$ follows (\ref{ultimate}) and $|\mathcal{N}{(i)}|$ is the cardinality of $\mathcal{N}{(i)}$. As we do not observe $X_k$ for $k \in \mathcal{M}$, here $Y_k^{mis}$ is imputed according to the proportion of disease spots in the neighboring spots of $k$.

When missing is nonignorable, we carry out inference on (\ref{eqn:inf_obs_data}) directly. To facilitate this, we extend (\ref{ultimate_fram}) to incorporate the missing mechanism of $R_i$ via (\ref{ultimate}) and (\ref{model:nonignorable_missing_model}), which gives
\begin{equation*}
\begin{aligned} \label{methodology: nonignorable_missing_bayesian_framework}
Y_i^{obs}| X_i, Y_j^{obs}, {Y_j}^{mis}, R_j \stackrel{i.i.d}{\sim} \text{Bernouli}(\mu_i),  \ \text{for} \ i=1,\dots,n, j \in \mathcal{N}(i),
\\{Y_k}^{mis} \sim \text{Bernouli}\left(\frac{1}{{|\mathcal{N}{(k)}|}}\sum_{w \in \mathcal{N}(k)}\left\{Y_w^{obs} R_w+{Y_w}^{mis} (1-R_w)\right\}\right), \ \text{for} \ k \in \mathcal{M},
\\  R_i|Y_i, R_j \sim \text{Bernouli} \left\{\text{P} (R_i=1)\right\}, \ \text{for} \ i=1,...,n, \ j \in \mathcal{N}(i),
\end{aligned}
\end{equation*}

\noindent where $P(R_i=1)$ is specified in (\ref{model:nonignorable_missing_model}).
\section{Parameter Estimation Procedure}\label{sec:estimation}

\subsection{No-U-Turn Sampling} 
We adopt the No-U-Turn Sampler \citep[NUTS,][]{r2}, a modified Hamiltonian Monte Carlo (HMC) algorithm commonly used for parameter estimation in Bayesian hierarchical model, particularly for absolutely continuous spike prior in (\ref{SSM}).

To describe the sampling algorithm, we first let $\theta_{t}$ be the parameters generated at the $t$th state of the Markov chain from our target distribution of $\theta$, and let $p_t$ to be the auxiliary Gaussian momentum variables generated at the $t$th step, which is assumed to follow the distribution $N(0,M)$ independent of $t$ and $\theta$, with $M$ being a covariance matrix prespecified. We further define $H(\theta_t,p_t) = -\text{log}f(\theta_t) + \frac{1}{2}p_t^{\intercal}M^{-1}p_t$ with $\theta_t$ and $p_t$ independent for every $t$. The joint probability density function of $\theta_t$ and $p_t$ can be derived as

\begin{equation*}
    \begin{aligned}\label{HD2}
        f(\theta_t,p_t)=\text{exp}\left\{\text{log} f(\theta_t)+\text{log} f(p_t)\right\} \propto \text{exp}\left\{\text{log} f(\theta_t)-\frac{1}{2}p_t^{\intercal}M^{-1}p_t \right\} \propto \text{exp}\left\{-H(\theta_t,p_t)\right\},
    \end{aligned}
\end{equation*}

\noindent where $H(\theta_t,p_t)$ is often known as Hamiltonian equation, and $f(\theta_t)$ is the probability density function of  $\theta_t$. Then, $\theta_t$ is sampled by the NUTS algorithm via a three-step procedure. 
\begin{itemize}

\item \textbf{Step 1}: given initial values of $\theta_{t-1}$ and $p_{t-1}$, we apply slice sampling  \citep{r3} by generating an auxiliary variable $u_t$ with
\begin{equation}\label{GFU}
    \begin{aligned}
         p(u_t|\theta_{t-1},p_{t-1}) \sim \text{Uniform}[0,\text{exp}\left\{-H(\theta_{t-1},p_{t-1})\right\}],
    \end{aligned}
\end{equation}

\noindent such that the joint distribution of $u_t, \theta_{t-1}, p_{t-1}$ is given by

\begin{equation*}
  f(u_t,\theta_{t-1},p_{t-1})=\left\{
  \begin{aligned}
  \frac{1}{C}, \ 0\leq u_t \leq \text{exp}\left\{-H(\theta_{t-1},p_{t-1})\right\},
  \\0,\text{otherwise},
\end{aligned}\right.
\end{equation*}

\noindent where $C$=$\int \text{exp}\left\{-H(\theta_{t-1},p_{t-1})\right\} d\theta_{t-1} dp_{t-1}$ is a normalizing constant.

\item \textbf{Step 2}: the sampling of ($\theta_t$, $p_t$) is then driven by sampling $(u_t,\theta_{t-1}, p_{t-1})$ alternatively from the joint density distribution $f(u_t,\theta_{t-1},p_{t-1})$. 

For a $u_t$ obtained from Step 1, we adopt doubling method \citep{r3} to sample a series of candidates $B_t$=$\left\{(\theta^{*1},p^{*1}),\dots,(\theta^{*m},p^{*m})\right\}$ from
\begin{equation*}
    \begin{aligned}
        (\theta^{*i},p^{*i}) \in S= \left\{u_t\leq \text{exp}(-H(\theta^{*i},p^{*i}) )\right\}, \text{for} \ i\in \left\{1,\dots,m\right\}.
    \end{aligned}
\end{equation*}

With invariant property of $H(\theta,p)$ \citep{r4}, the doubling method proceeds randomly taking either forward or backward leapfrog formula to sample $(\theta^{*i},p^{*i})$, for $i=1,\dots, m$.  Specifically, we start with initial values $\tau = 0, i = 0$, and $\theta(\tau)=\theta_{t-1}$. Then we iteratively:
\begin{enumerate}

\item generate momentum $p(\tau)=p_t$ from the distribution $N(0,M)$;
 
\item generate $v_i$ uniformly from $\left\{-1,1\right\}$;
\item conduct a half step for $p_t$ with
\begin{equation}\label{NUTS_step_1}
    \begin{aligned}
       p(\tau+\frac{1}{2} v_i \epsilon)=p(\tau)-\frac{1}{2} v_i \epsilon\frac{\partial f(\theta)}{\partial \theta}(\theta(\tau)),
    \end{aligned}
\end{equation}
\noindent where $\epsilon$ is the integration step size pre-determined;
\item perform the second half step for $\theta$ and $p$ respectively,
\begin{equation}\label{NUTS_step_2}
    \begin{aligned}
 \theta(\tau+ v_i\epsilon)=\theta(\tau)+ v_i \epsilon p(\tau+\frac{1}{2} v_i \epsilon),
\end{aligned}
\end{equation}

and
\begin{equation}\label{NUTS_step_3}
    \begin{aligned}
        p(\tau+ v_i \epsilon)=p(\tau+\frac{1}{2} v_i \epsilon)-\frac{1}{2} v_i \epsilon\frac{\partial f(\theta)}{\partial \theta}\left\{\theta(\tau + v_i \epsilon)\right\};
    \end{aligned}
\end{equation}
\item the above (\ref{NUTS_step_1}), (\ref{NUTS_step_2}), and (\ref{NUTS_step_3}) complete one full leapfrog step for $\theta$ and $p$, and we assign $\theta^{*i}=\theta(\tau+ v_i\epsilon)$,and $p^{*i}= p(\tau+ v_i \epsilon)$;
\item repeat steps 2) - 5) for $2^i$ times;
\item iterate $i=i+1$, doubling proceeds will stop when 
\begin{equation*}
    \begin{aligned}
        (\theta^{+}-\theta^{-})^{\intercal}p^{-}<0 \ \text{or} \ (\theta^{-}-\theta^{+})^{\intercal}p^{+}<0,
    \end{aligned}
\end{equation*}

\noindent where $(\theta^{+},p^{+})$ and $(\theta^{-},p^{-})$ are respectively $(\max_{j\leq i} \theta^{*j}, \max_{j\leq i} p^{*j})$, $(\min_{j\leq i} \theta^{*j}, \min_{j\leq i} p^{*j})$.

\end{enumerate}

 uniformly sample $(\theta^{*},p^{*})$ from $B_t$ and accept it as ($\theta_t, p_t$)=($\theta^{*}, p^{*}$).

\end{itemize}

In practice, NUTS can be implemented by package \textbf{Rstan} in \textsf{R}.

\subsection{Automatic Differentiation Variational Inference}

While the NUTS algorithm efficiently samples the parameters from their posterior distributions, its scalability remains a challenge, especially in the context of large-scale and high-dimensional datasets as commonly encountered in spatial transcriptomics. To address the high computational demands of our model, we adopt Automatic Differentiation Variational Inference (ADVI), which was proposed by \cite{kucukelbir2017automatic}. This method serves as a more computationally efficient alternative by approximating the posterior distribution of the parameters, taking the tradeoff between computational feasibility and the accuracy of the estimated parameters.

\subsubsection{Transformation of constrained variables}\label{subs:trans.cons.var}

ADVI requires the support of each parameter to be $\mathbb{R}$. To facilitate this, we first map the support of $\theta = (\eta,\beta,\iota,\tau,w)^{\intercal}$ into $\mathbb{R}^{p_\theta}$, where $p_\theta$ is the dimension of $\theta$, by taking the transformations $T(\theta)=(\Phi^{-1}(\frac{\eta}{c_1}),\beta,log(\tau^2),\Phi^{-1}(w))$, and let the resulting transformed parameters be $\zeta=T(\theta)$. The transformed joint density $f(\zeta,y|x)$ has the representation, $f(\zeta,y|x)=f(T^{-1}(\zeta),y|x)|\text{det} \  J_{T^{-1}}(\zeta)|=f(y|x,T^{-1}(\zeta))f(T^{-1}(\zeta))|\text{det} \  J_{T^{-1}}(\zeta)|$, where the detailed form of $f(\zeta,y|x)$ in Appendix is indicated on (A.1).

\subsubsection{Full-rank Gaussian Variational Approximation}
The computation of $f(\zeta,y|x)$ can be challenging due to its complex forms, we consider a full-rank Gaussian variational approximation
\begin{equation}\label{VI: full-rank Gaussian VI}
    \begin{aligned}
        q(\zeta; \phi)=\text{Normal}(\zeta|\mu_q,\Sigma_q)
    \end{aligned}
\end{equation}

\noindent for $f(\zeta,y|x)$ in (A.1) where $\phi=(\mu_q,\Sigma_q)$. In order to ensure the $\Sigma_q$ to be positive semidefinite, we reparameterize the covariance matrix using a Cholesky factorization, $\Sigma_q=LL^{\intercal}$. Then the estimation of parameters $\mu_q$ and $\Sigma_q$ from (\ref{VI: full-rank Gaussian VI}) can be obtained by maximizing the evidence lower bound (ELBO), which is given by
\begin{equation}
    \begin{aligned}\label{eqn:ADVI_ELBO}
        \mathcal{L}=\text{E}_{\zeta\sim q(\zeta)}\left\{\text{log} f(y,T^{-1}(\zeta)|x)+log |\text{det} J _{T^{-1}}(\zeta)|\right\}+\text{H}\left\{q(\zeta;\phi)\right\},
    \end{aligned}
\end{equation}

\noindent where the expectation is taken with respect to $\zeta$, and $\text{H}\left\{q(\zeta;\phi)\right\}=-\int q(\zeta;\phi)\text{log}\left\{q(\zeta;\phi)\right\}d\zeta$.

\subsubsection{Stochastic optimization}
In practice, to further improve the computational efficiency, the optimization of (\ref{eqn:ADVI_ELBO}) is facilitated by stochastic optimization. Specifically, 
We first perform an elliptical standardization transformation, $\xi=L^{-1}(\zeta-\mu_q)$ to convert the Gaussian variational approximation into a standard Gaussian. This transformation results in a variational density free of parameters,
\begin{equation*}
    \begin{aligned}
        q_0(\xi)=\text{Normal} (\xi|0,\mathbb{I}),
    \end{aligned}
\end{equation*}

\noindent where $\mathbb{I}$ is the identity matrix, and then the first term on the right hand of (\ref{eqn:ADVI_ELBO}) becomes
\begin{equation*}
    \begin{aligned}\label{Exp about SGD}
        \mathcal{L}(\xi)=\text{E}_{q \sim q_0}[\log \ f(Y,T^{-1}(\mu_q+L\xi))+\log |\text{det} J _{T^{-1}}(\mu_q+L\xi)|].
    \end{aligned}
\end{equation*}

\noindent Here, the expectation is taken with respect to a standard Gaussian probability density function. Then, stochastic gradient ascent is employed to maximize (\ref{eqn:ADVI_ELBO}), and we will have,

\begin{equation}
    \begin{aligned}\label{advi_mu}
        \diffp*{\mathcal{L}}{\mu_q}=\text{E}\left[\frac{\partial}{\partial \zeta}{\text{log} f(Y,T^{-1}(\zeta))} \frac{\partial}{\partial \zeta}T^{-1}(\zeta)+ \frac{\partial}{\partial \zeta} \text{log} |\text{det}J_{T^{-1}}(\zeta)| \right],
    \end{aligned}
\end{equation}

\begin{equation}
    \begin{aligned}\label{advi_L}
        \diffp*{\mathcal{L}}{L}=\text{E}\left[\left\{\frac{\partial}{\partial \zeta} \text{log} f(Y,T^{-1}(\zeta))\frac{\partial}{\partial \zeta}T^{-1}(\zeta)+\frac{\partial}{\partial \zeta} \text{log} |\text{det}J_{T^{-1}}(\zeta)|\right\} \xi^{\intercal}\right]+(L^{-1})^\intercal.
    \end{aligned}
\end{equation}

 The gradients inside the expectation can be calculated with automatic numerical differentiation. The intractable expectation is computed by Monte Carlo integration. That is, we draw samples from standard multivariate Gaussian distribution and evaluate the empirical mean of the gradients within the expectation. The detailed procedure is given in Algorithm~1. In practice, we can implement ADVI using \textbf{Rstan} based on the pseudo-code given in Algorithm \ref{alg1}.

\begin{algorithm}
\caption{Automatic Differentiation Variational Inference (ADVI)}
\begin{algorithmic} \label{alg1}
    \STATE Set iteration number to $i=1$
    \STATE Initialize $\mu^1=0$
    \STATE Initialize $\mathbf{L^1=}$ $\text{I}$
    
    \WHILE{$\text{ELBO above threshold}$}
    \STATE Draw $\xi$ from the Standard Multivariate Normal Distribution
    \STATE Evaluate $\diffp*{\mathcal{L}}{\mu_q}$ using equation (\ref{advi_mu})
    \STATE Evaluate $\diffp*{\mathcal{L}}{L}$ using equation (\ref{advi_L})
    \STATE state stepsize $s^{i}$ (a pre-specified value)
    \STATE $\mu^{i+1}\leftarrow\mu^{i}+\text{diag}(s^{i})\diffp*{\mathcal{L}}{\mu_q}$
    \STATE $L^{i+1}\leftarrow L^{i}+\text{diag}(s^i)\diffp*{\mathcal{L}}{L}$
    \STATE Increment iteration number
    \ENDWHILE
    \STATE Return $\hat{\mu}\leftarrow \mu^{i}$
    \STATE Return $\hat{L}\leftarrow L^{i}$
\end{algorithmic} 
\end{algorithm}

\section{Simulation Studies }\label{sec:Simulation}

We conduct simulation studies to assess the performance of the proposed model in parameter estimation. We compare the autologistic model (\ref{restriautomodel}) with the naive approach, where the spatial patterns are disregarded. We examine three distinct scenarios: in Scenario 1, we consider all the spots are displayed on a single slice; in Scenario 2, we consider that a more complex data structure with the spots may come from different slices; and the third scenario includes the situation where missing data present.

 For each estimator, we report the average bias (denoted “avgBias”), the average empirical standard error (denoted “avgSEE”), the average model standard error (denoted “avgSEM”), the average coverage rate (denoted “avgCR”) for 95 percent credible intervals of the model parameters among 200 simulation iterations. 
 
\subsection{ Simulation 1: a Single Slice Scenario} \label{Sim1}

\begin{figure}
\centering
\begin{subfigure}{0.4\textwidth}
    \includegraphics[width=\textwidth]{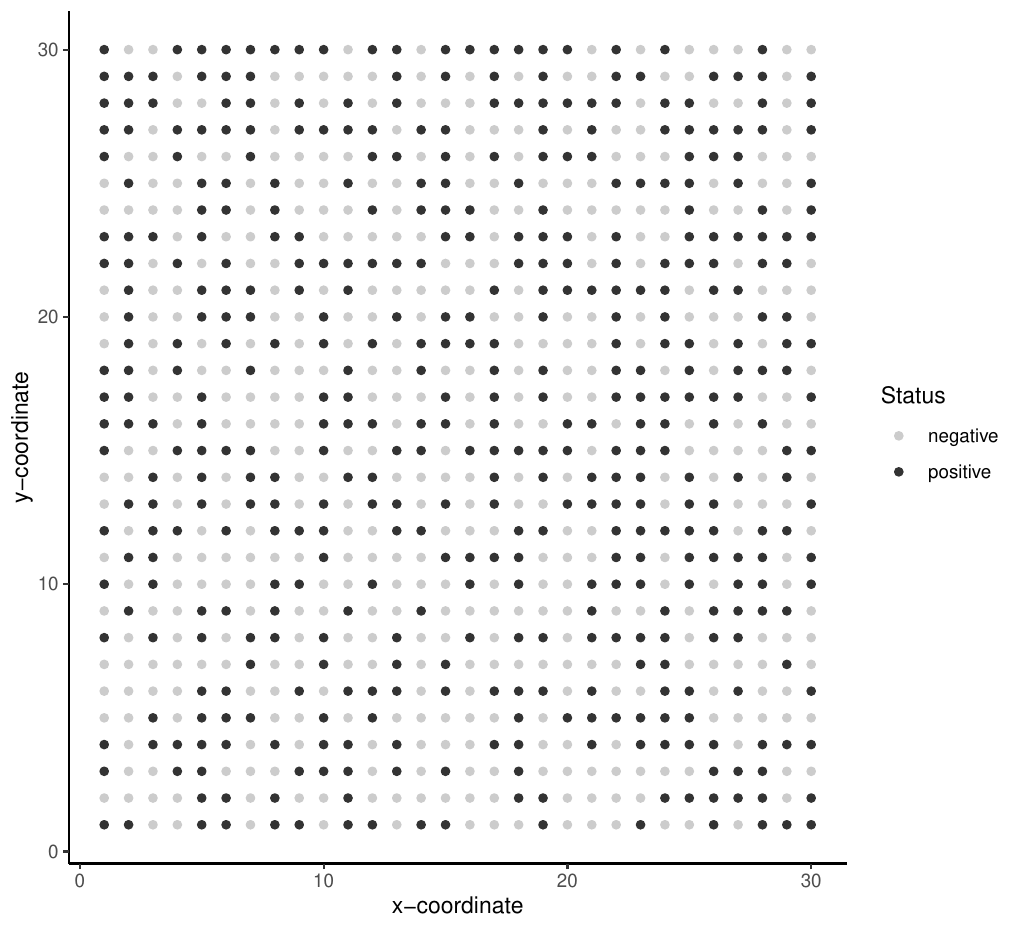}
\end{subfigure}
\hfill
\begin{subfigure}{0.4\textwidth}
    \includegraphics[width=\textwidth]{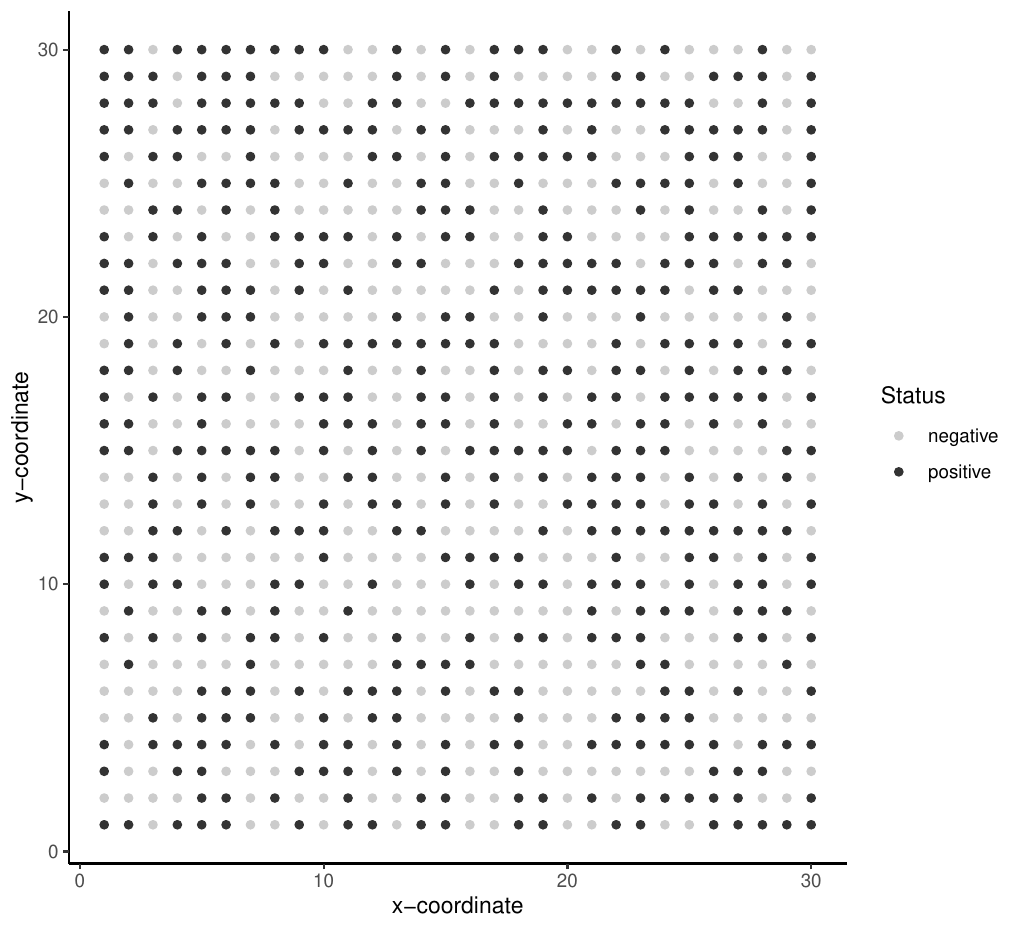}
\end{subfigure}
\hfill
\begin{subfigure}{0.4\textwidth}
    \includegraphics[width=\textwidth]{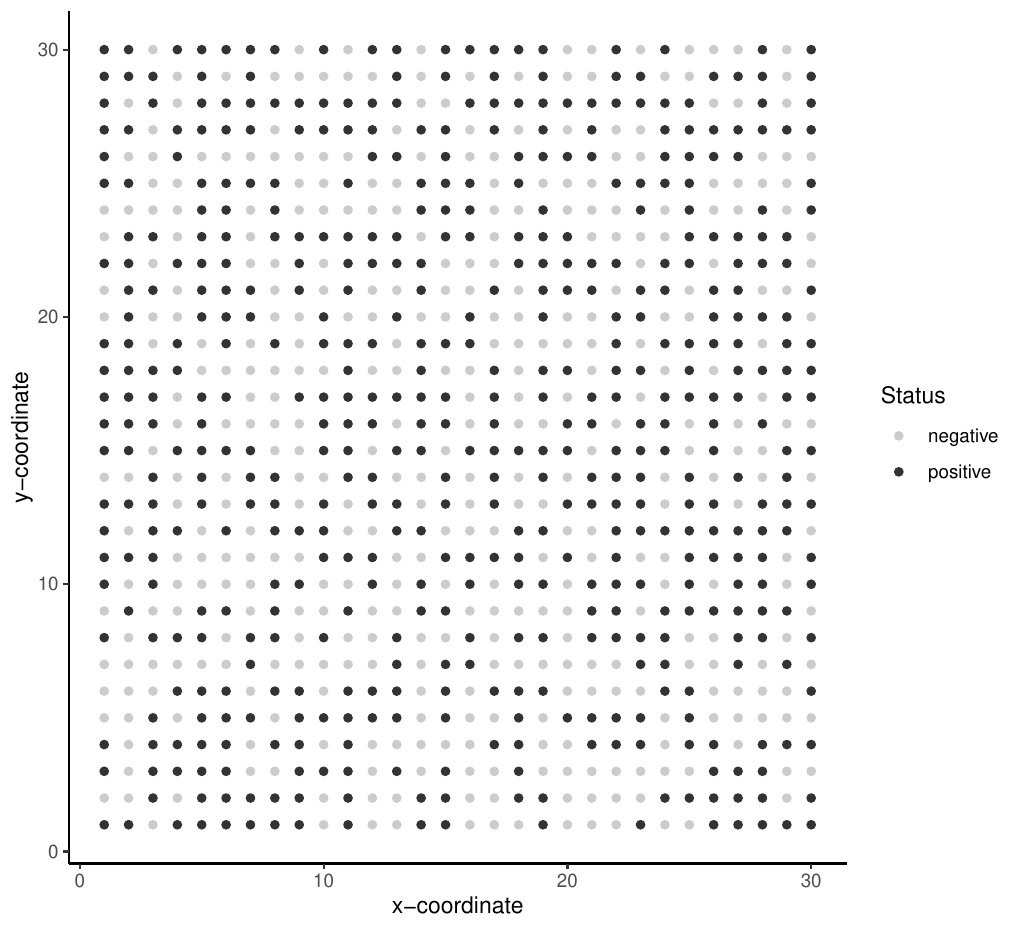}
\end{subfigure}
        
\caption{Exemplary figures of the generated disease outcomes in a 30$\times$30 lattice in Simulation 1, under the scenarios of $\eta=0.4, 1.6, 2.8$.}
\label{fig:sim1_lattice}
\end{figure}

 In this simulation study, we consider the sample spots are displayed on a $30\times30$ lattice and for each spot (indexed as $i$), we generate 20 continuous covariates $X_{ij}$, independently from a standard normal distribution, for $j=1,...,20$. A Gibbs sampler is used to simulate $Y_i$ with the parameters $\beta = (1,2,3,-4,-5,\widetilde{0}_{15})^{\intercal}$, where $\widetilde{0}_{15}$ is a 15-dimensional zero vector. For each simulation, a random starting lattice is generated, followed by the execution of a total of 2000 iterations of the Gibbs sampler. The last iteration was taken as the realization of the lattice.  Consistent with \cite{hughes2011autologistic}, we evaluate model (\ref{restriautomodel}) with varying autocorrelation factors $\eta$ = 0.4, 1.6, and 2.8, representing low, medium, and high autocorrelation, respectively \citep{sun2008bayesian}. We set $c_{1}$ in (\ref{SSM}) to be 8. Exemplary simulated outcomes on the lattices are displayed in Figures \ref{fig:sim1_lattice}. Furthermore, we take $b_1=5, b_2=50$, and $v_0 = 0.000001$, a small value close to 0, to provide a weakly informative prior for $\tau$, which is consistent with \cite{r1}.

\begin{figure}[h]
\centering
\includegraphics[width=0.8\textwidth]{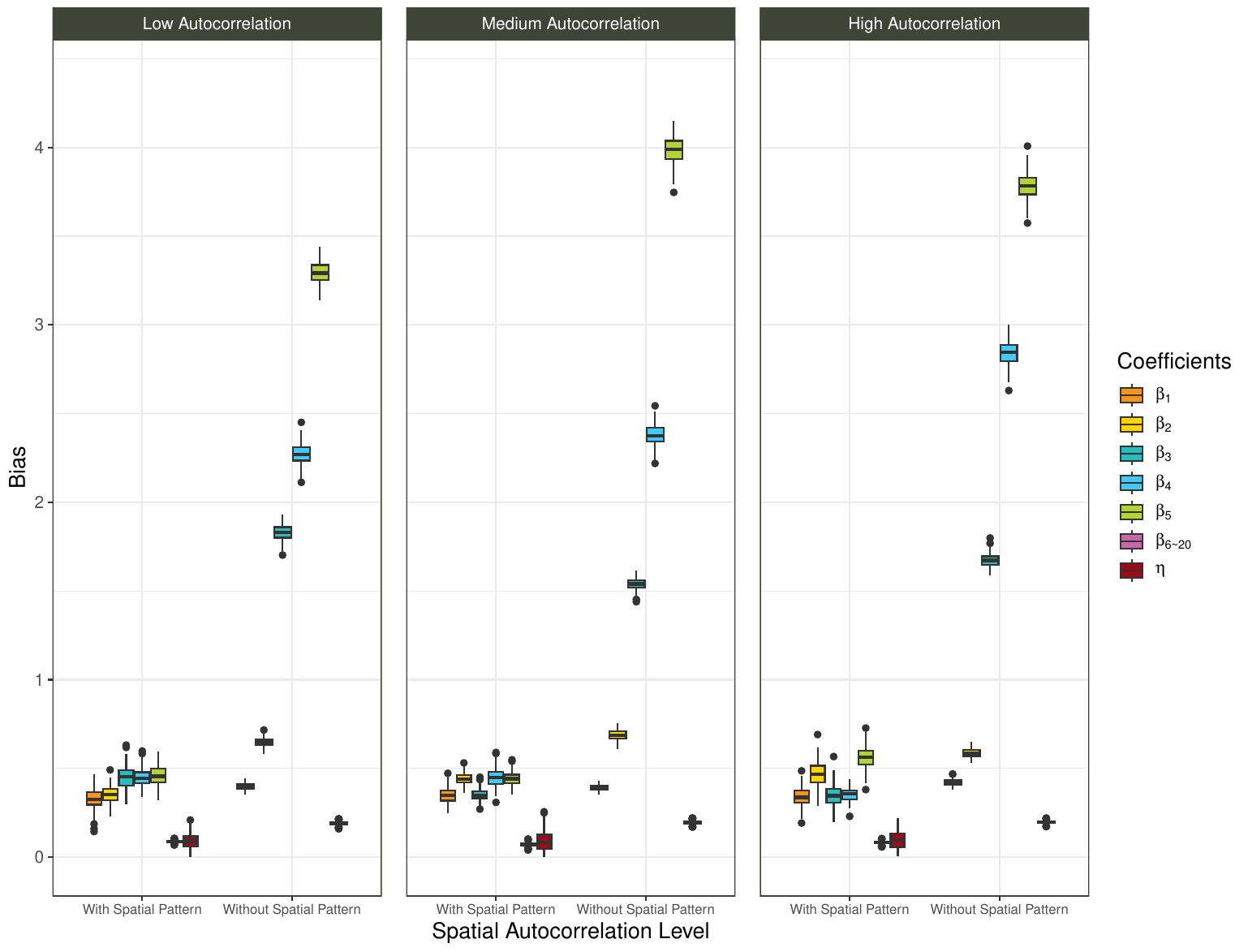}
\caption{Boxplot of the biases for the parameter estimation in Simulation 1 implemented by ADVI algorithm. The parameters $\beta_1$ to $\beta_5$ have non-zero true values, while $\beta_{6-20}$ represent the remaining parameters with true values of zero, $\eta$ indicates the parameter of spatial correlation, and their boxes show their distributions of bias. }
\label{fig:sim1_vi}
\end{figure}

\begin{figure}[h]
\centering
\includegraphics[width=0.8\textwidth]{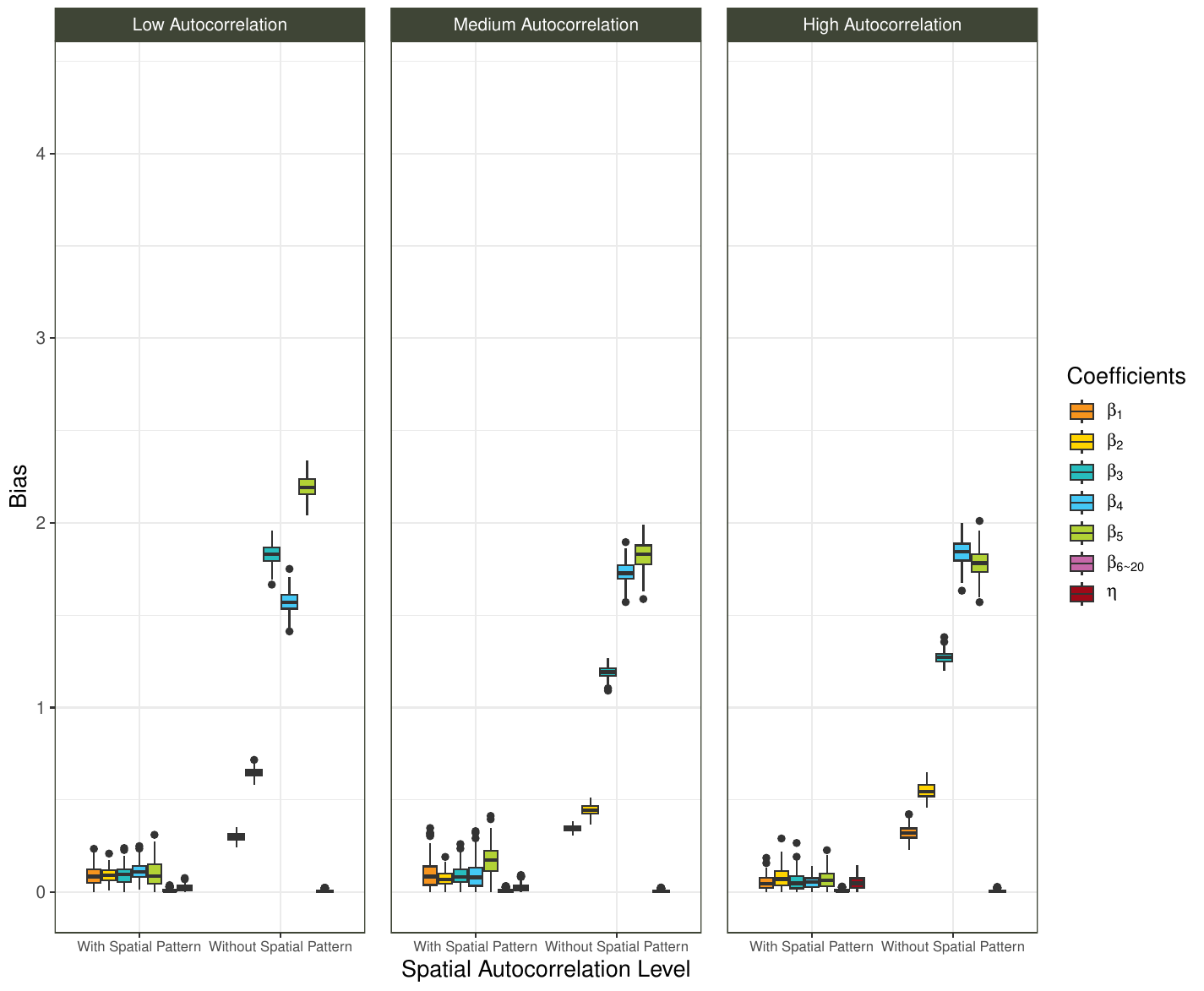}
\caption{Boxplot of the biases for the parameter estimation in Simulation 1 implemented by NUTS algorithm. The parameters $\beta_1$ to $\beta_5$ have non-zero true values, while $\beta_{6-20}$ represent the remaining parameters with true values of zero, $\eta$ indicates the parameter of spatial correlation, and their boxes show their distributions of bias.}
\label{fig:sim1_nuts}
\end{figure}

We compare the performance of the proposed autologistic model (\ref{restriautomodel}) implemented by both NUTS and ADVI to the naive model where spatial correlations are ignored. The results for bias are presented in Figures \ref{fig:sim1_vi} and \ref{fig:sim1_nuts}, the detailed results in “avgBias”, “avgSEE”, “avgSEM” and “avgCR” are shown in Tables \ref{tab:sim1_1}-\ref{tab:sim1_3} in Appendix. It can be seen that the naive model generally produces a larger bias in the parameters. On the contrary, the proposed model adjusts for the spatial effects, consistently resulting in a reduced average bias for parameter estimations regardless of the degree of the spatial correlation.

\begin{figure}[h]
\centering
\includegraphics[width=0.8\textwidth]{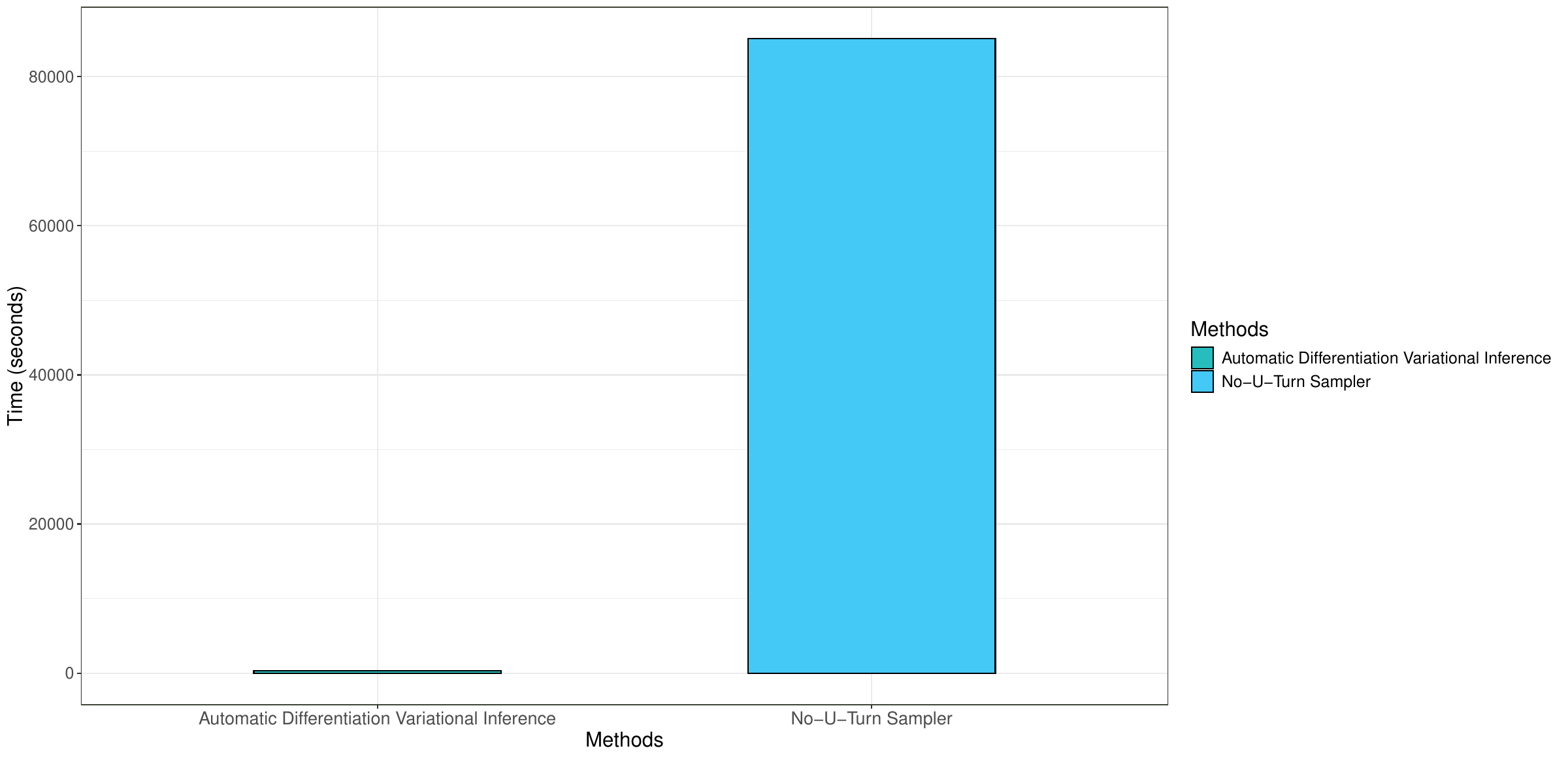}
\caption{Comparison of running time (in seconds) for the parameter estimation of (\ref{restriautomodel}) implemented by ADVI and NUTS, respectively.}
\label{fig:sim1_time_compar}
\end{figure}

 It was noteworthy that the biases of the sampling method are generally lower than that of variational inference in this context. Both methods demonstrate reasonable performance in estimating standard error in various settings, maintaining a valid coverage rate. However, the variational inference method demonstrates a much higher computational efficiency as reflected by the computing time. As shown in Figure \ref{fig:sim1_time_compar}, while NUTS necessitates 23 hours to execute a single simulation study, ADVI requires mere 5 minutes. Consequently, in the following simulations and data analysis on a larger scale data set, ADVI will be exclusively employed for implementation.

\subsection{Simulation 2: Evaluation of Model with Multiple Slices}

In this simulation study, we validate the efficacy of our proposed model in handling multiple slices simultaneously, as delineated in Section \ref{model_for_multiple_slice}. For each slice, we generate  $X_{gij}$ independently in the same way as in Section \ref{Sim1}, and we consider the number of slices $G=6$. Same as in Section \ref{Sim1}, a Gibbs sampler is employed to generate $Y_{gi}$ for $g=1,\dots,6$, and $i=1,\dots,900$, following (\ref{eqn:general_equation_ms}) with  $\eta=1.6$, $\beta$ to be the same as in Section \ref{Sim1}, and $U_{cg}$ are generated from $N(0,\Sigma_c)$ where $\left\{\sigma^2, \rho \right\}$ are chosen to be $\left\{0.1,0.1        \right\}$, $\left\{0.1,0.4        \right\}$, $\left\{0.4,0.1        \right\}$, $\left\{0.1,0.4        \right\}$, respectively, which reflects the different levels of variance and correlation among slices. A total of 200 simulation studies were conducted by ADVI, and we consider the same values of $b_1, b_2$, and $v_0$ as in Section \ref{Sim1}.

\begin{figure}[h]
\centering
\includegraphics[scale=0.4]{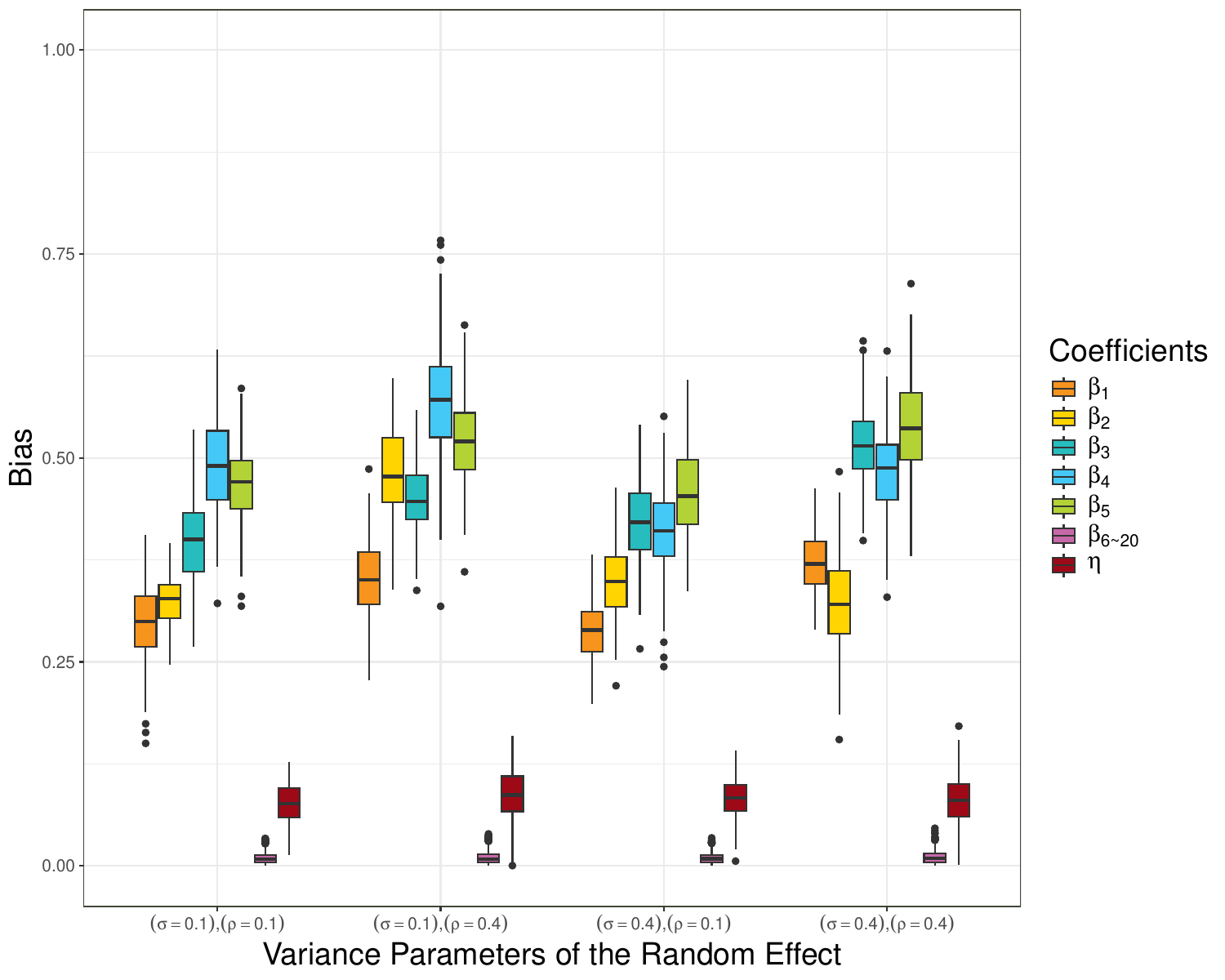}
\caption{Boxplot of the biases for parameter estimation in Simulation 2 with varying choices of $\rho$ and $\sigma$. The parameters $\beta_1$ to $\beta_5$ have non-zero true values, while $\beta_{6-20}$ represent the remaining parameters with true values of zero, $\eta$ indicates the parameter of spatial correlation, and their boxes show their distributions of bias.}
\label{fig:Sim3}
\end{figure}

The results for bias are reported in Figure \ref{fig:Sim3} and Table \ref{tab:sim3_bias}. It can be seen that our model performs reasonably in both point estimation and variation estimation regardless of different levels of variations and correlations among slices.

\subsection{Simulation 3: Evaluation of Model with Missing Data}

In this subsection, we evaluate the performance of the proposed method, where missing data are presented, described in Section \ref{ModelMiss}.

The covariates and true responses are generated in the same way as in Section \ref{Sim1}. We examine two simulation scenarios for introducing missing in data under the assumption of ignorable and nonignorable missing data mechanisms, respectively. To introduce in-sample missingness in the ignorable setting, we randomly mask spots within the lattice that are not on the periphery, setting their corresponding $R_i$ values to be 0. We considered the number of masked spots to be 10 and 30 respectively, reflecting varying degrees of missingness. For nonignorable missing mechanisms, Gibbs sampler is employed to generate $R_i$ according to model (\ref{model:nonignorable_missing_model}) with $(\gamma_0,\gamma_1,\gamma_2)^{\intercal}=(-6,1,4)^{\intercal}$ and $(\gamma_0,\gamma_1,\gamma_2)^{\intercal}=(-5,1,1.6)^{\intercal}$, which on average will result in 1.1\% and 3.7\% of the total 900 spots being missing, respectively, making the number of missing spots in the nonignorable setting comparable to that in the ignorable setting. A total of 200 simulation studies are conducted. 

\begin{figure}[h]
\centering
\includegraphics[scale=0.45]{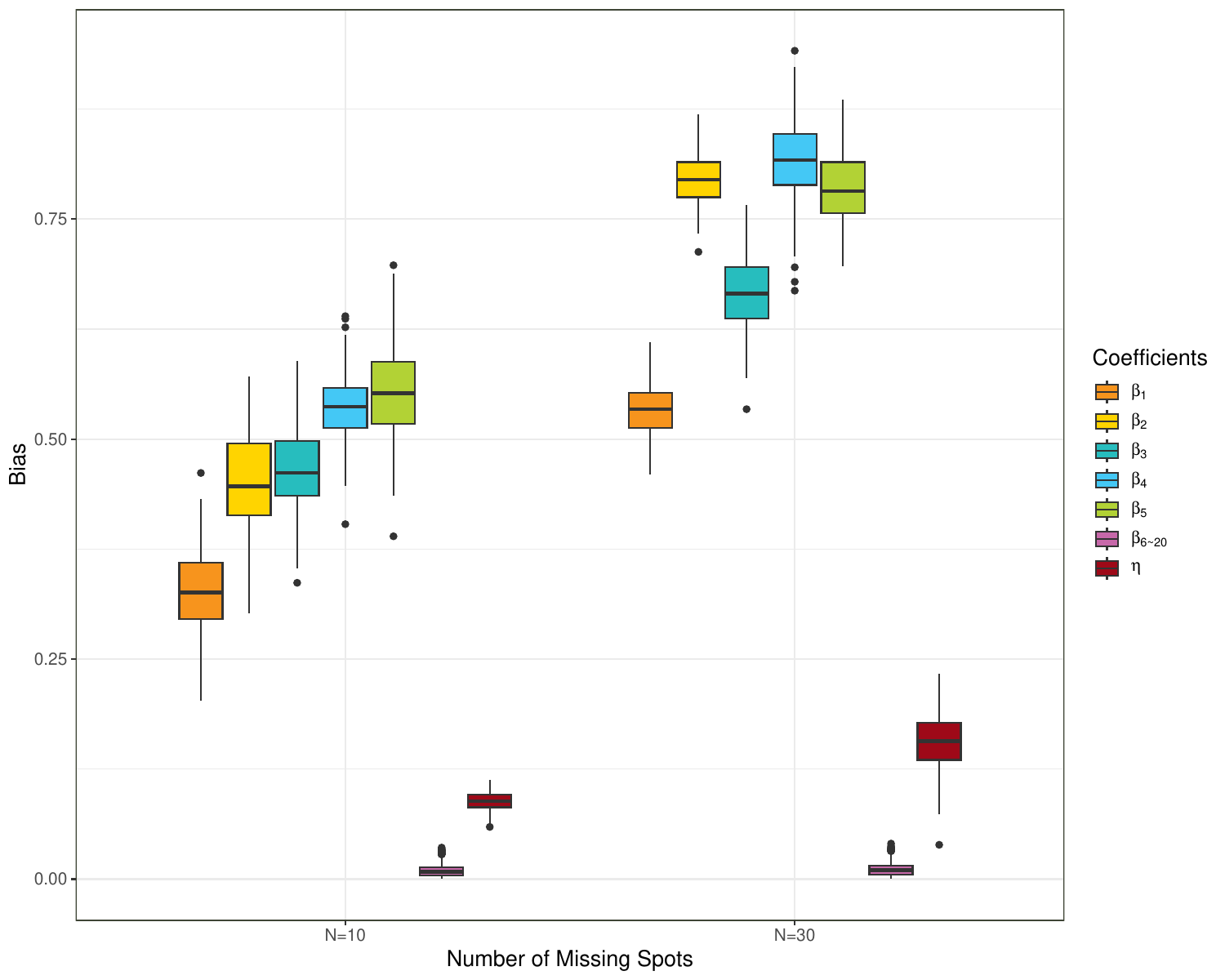}
\caption{Boxplot of the biases for parameter estimation in Simulation 3 with ignorable missing mechanism. The parameters $\beta_1$ to $\beta_5$ have non-zero true values, while $\beta_{6-20}$ represent the remaining parameters with true values of zero, $\eta$ indicates the parameter of spatial correlation, and their boxes show their distributions of bias.}
\label{fig:Sim2}
\end{figure}

\begin{figure}[h]
\centering
\includegraphics[scale=0.45]{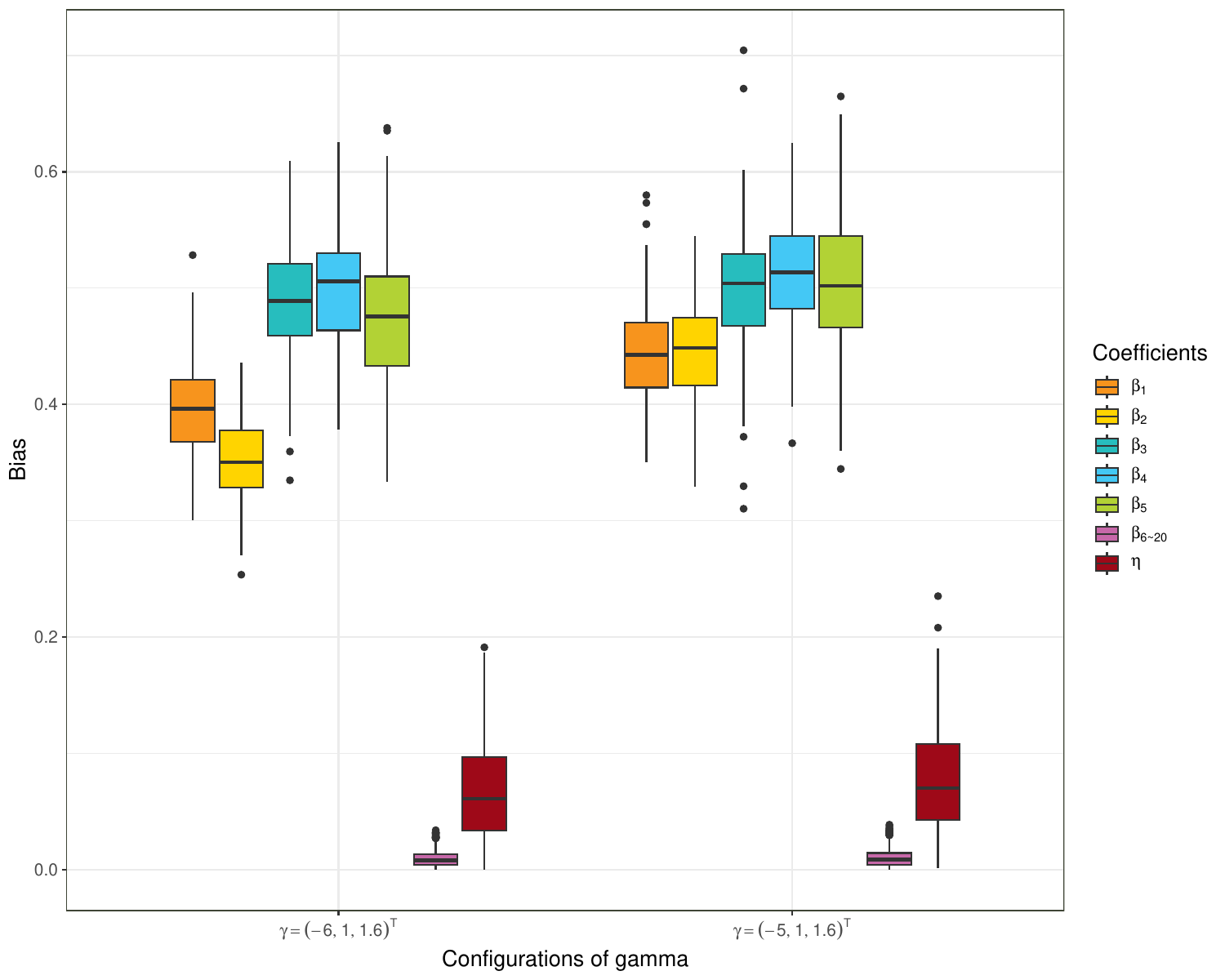}
\caption{Boxplot of the biases for parameter estimation in Simulation 3 with nonignorable missing mechanism. The parameters $\beta_1$ to $\beta_5$ have non-zero true values, while $\beta_{6-20}$ represent the remaining parameters with true values of zero, $\eta$ indicates the parameter of spatial correlation, and their boxes show their distributions of bias.}
\label{fig:Sim2_nonign}
\end{figure}

The results for ignorable setting is reported in Figure \ref{fig:Sim2}, Table \ref{sim3_1} and those for nonignorable setting are shown in Figure \ref{fig:Sim2_nonign} and Table \ref{sim3_2}. Comparing with the method outlined in Section \ref{bayesian1}, we observe that when the number of missing points is low (e.g., 10) in the ignorable missing setting, the method exhibits a similar capability to the one described in Section \ref{bayesian1}. It is seen that the bias of the model may be relevant to the number of missing spots as reflected when the number of missing points reaches 30. Nonetheless, both models demonstrate comparable performance to model (\ref{restriautomodel}) maintaining a valid coverage rate, and are far better estimation bias than the naive model. For the nonignorable missing mechanism, we observe similar patterns as found in the ignorable setting, demonstrating the performance of the proposed model in handling the complex missing mechanism.

\section{Analysis of HER2-positive Breast Cancer Spatial Transcriptomic Data }\label{sec:Application}

We illustrate the usage of the proposed method by applying it to a spatial transcriptomic dataset arising from a HER2-positive breast cancer study \citep{r5}. The datasets include gene expression measurements for over 15000 genes from eight individuals (labeled with patient ID A-H). Among them, patients A, B, C, D have six slices, and patients E, F, G, H have three slices. In each slice, on average of 346 spots are sampled with their 2D spatial location recorded. The disease statuses are manually annotated by the pathologists, and each sampled spot is labeled as either cancer or non-cancer. Although a large number of genes are profiled, only a small portion of them can be relevant to the disease outcome. Our objectives in the study are 1) to identify the genes that are specific to the disease using the proposed model; and 2) then construct the predictive model to inform the risk of cancer status on each spot for any newly collected spatial transcriptomic sample.

Before conducting the analysis, we first perform preprocessing procedures. A gene filtering procedure is conducted to exclude the genes whose total counts are below 300. As the number of filtered genes are different for each slice, we take a common subset to ensure that only mutual genes are kept for the slices from the same patient. The gene expressions, originally recorded as counts (e.g. denoted as $X$), undergo a $\text{log}(X+1)$ transformation and are converted into a continuous scale. We introduce $Y_i$ be a binary variable to denote the outcome according to its disease status, where “0” or “1” stand for whether spot $i$ is located in a non-cancerous or cancerous region, respectively. We use model (\ref{eqn:general_equation_ms}) described in Section \ref{model:mult_slice_model} to analyze the multiple slices jointly. To accommodate for missing data, we consider the missing mechanism to be ignorable because the major cause of the missingness in this dataset is the tissue slicing procedure and no experimental biases are observed or recorded. Hence, we apply the strategy for handling ignorable missing in Section \ref{model:missing_data_model}. The specification of hyperparameters are displayed in Table \ref{real.data.hyperpara}. 
 
\begin{figure}[h]
\centering
\includegraphics[scale=0.4]{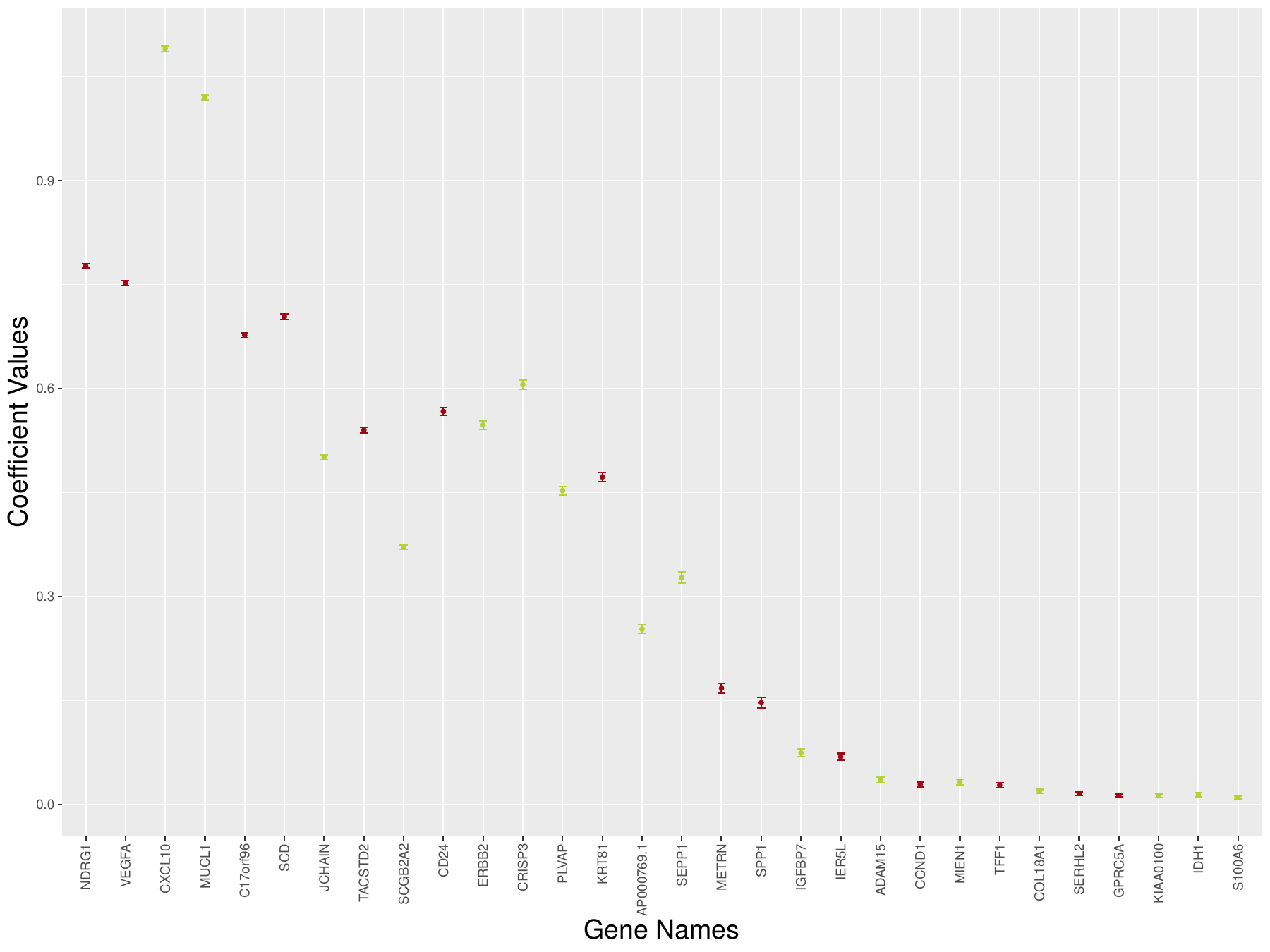}
\caption{Coefficients and confidence intervals for the top 30 genes in absolute values. The red and green colors indicate the gene is upregulated (positive $\beta$) or deregulated (negative $\beta$).}
\label{fig:da_50_genes_pval}
\end{figure}

\begin{figure}[h]
    \centering
    \begin{subfigure}{0.3\textwidth}
        \centering
        \includegraphics[width=\linewidth]{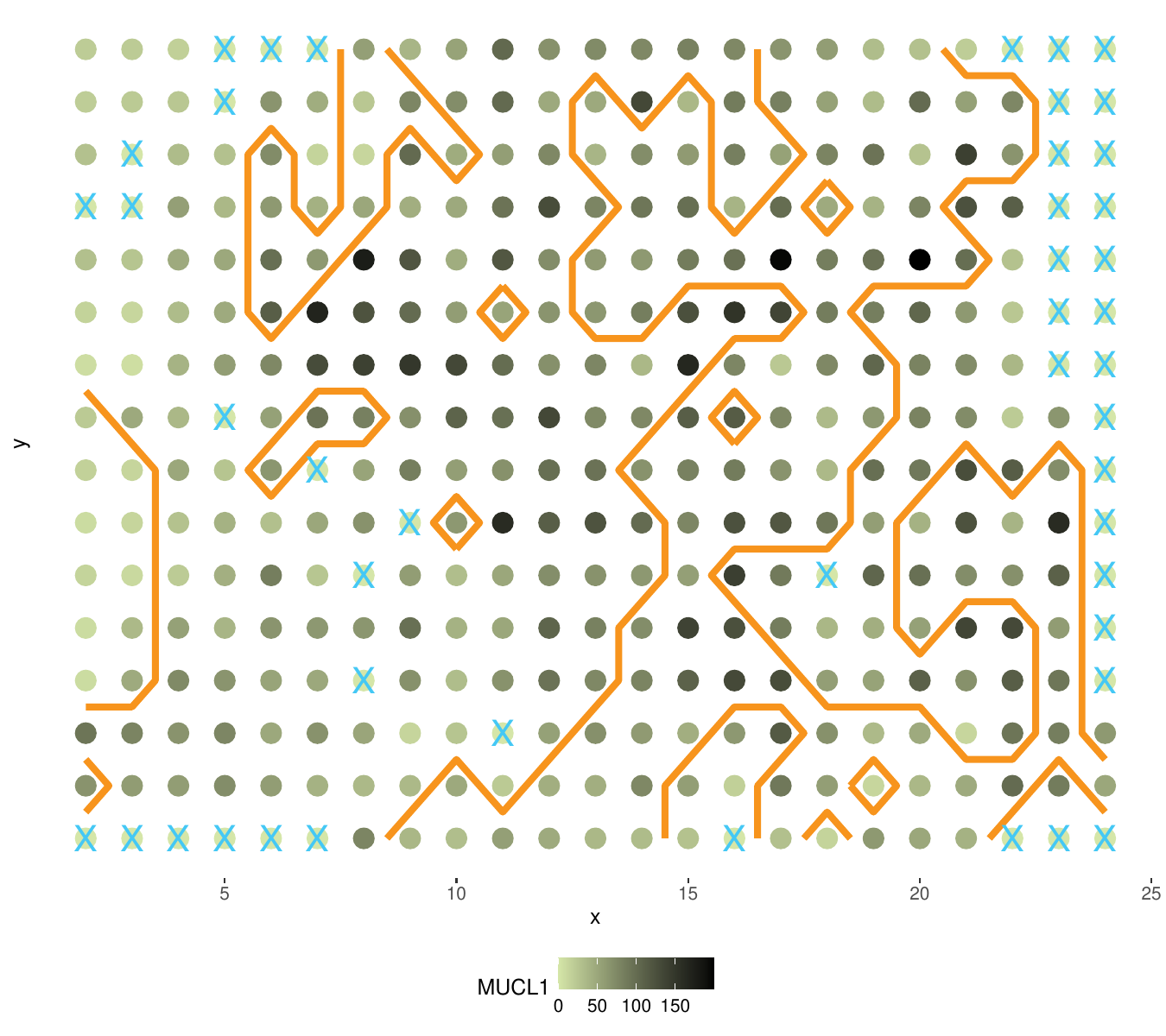}
        \caption{MUCL1}
        \label{fig:subfig1}
    \end{subfigure}
    \hfill
    \begin{subfigure}{0.3\textwidth}
        \centering
        \includegraphics[width=\linewidth]{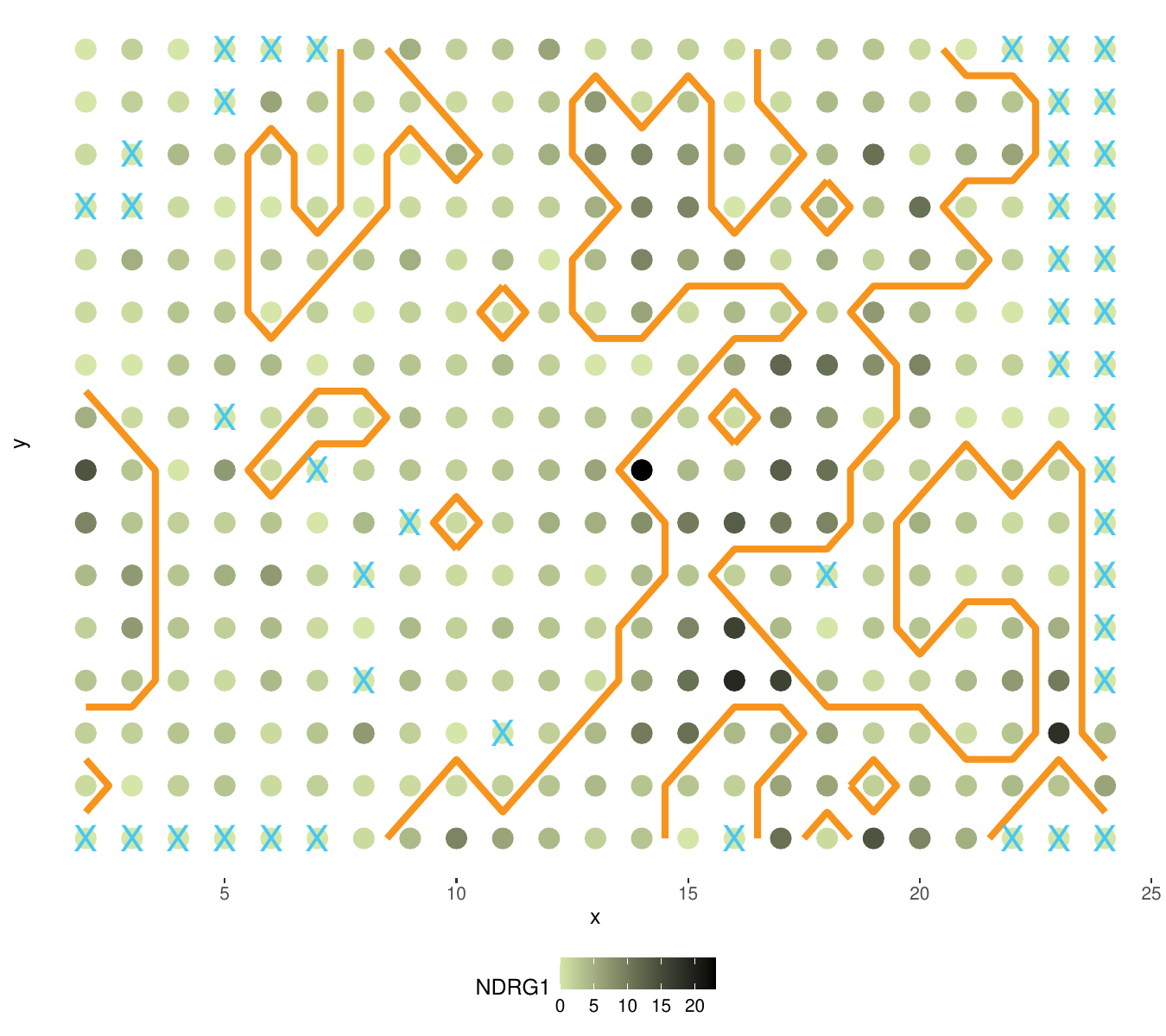}
        \caption{NDRG1}
        \label{fig:subfig2}
    \end{subfigure}
    \hfill
    \begin{subfigure}{0.3\textwidth}
        \centering
        \includegraphics[width=\linewidth]{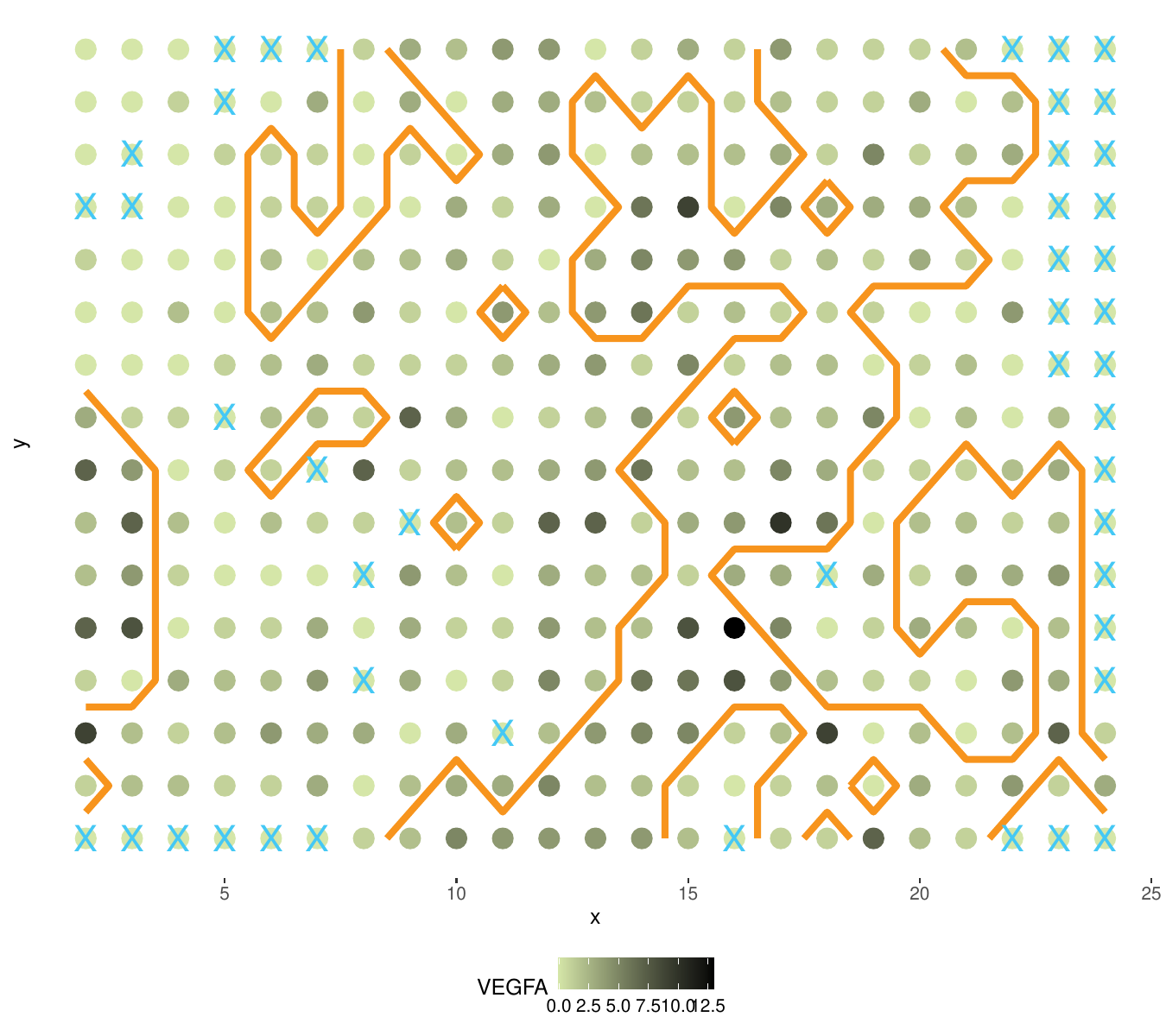}
        \caption{VEFGA}
        \label{fig:subfig3}
    \end{subfigure}
    \caption{The spatial patterns of gene expressions for MUCL1, NDRG1 and VEFGA. The darkness of the dot color implies the corresponding gene expression level on that spot. The orange contour reflects the boundary of the cancerous region and non-cancerous region. The blue cross (x) represents the missing spots.}
    \label{fig:da_countour}
\end{figure}

 Following the initial model fitting process, we identify the 30 genes that demonstrate the highest levels of effect to standardized deviation ratio, as we show in Figure \ref{fig:da_50_genes_pval}. To closely examine how the expressions of these genes are associated with the spatial pattern of both cancerous and non-cancerous regions within the tissue, the spatial visualization of gene expressions is generated in Figure \ref{fig:da_countour}. For example, the expression level of MCUL1 ($\hat{\beta}=-1.020$, 95\% CrI = $[-1.024,-1.016]$) is elevated mainly in non-cancerous regions and rarely in cancerous regions, suggesting its potential role in inhibiting cancer development. The finding is consistent with the results in Figure \ref{fig:da_50_genes_pval} that the estimated coefficient for the MUCL1 gene is negative. Similarly, as depicted in Figure \ref{fig:da_50_genes_pval}, the estimated coefficients for the NDRG1 ($\hat{\beta}=0.777$, 95\% CrI = $[0.774,0.780]$) and VEGFA ($\hat{\beta}=0.752$, 95\% CrI = $[0.749,0.756]$) genes are positive, indicating a likely promotion of cancer development. This is further reflected by Figure \ref{fig:da_countour} that the genes are highly expressed mostly in cancerous regions. Interestingly, despite this, the elevated expression locations for these two genes differ, suggesting distinct roles in cancer progression. These findings further support the necessity of modeling disease status by considering multiple gene expressions jointly. In addition, the estimated $\eta$ is 2.936, 95\% CrI=$[2.926,2.946]$, representing a notable level of spatial autocorrelation among adjacent spots.

\begin{figure}[h]
\centering
\includegraphics[scale=0.4]{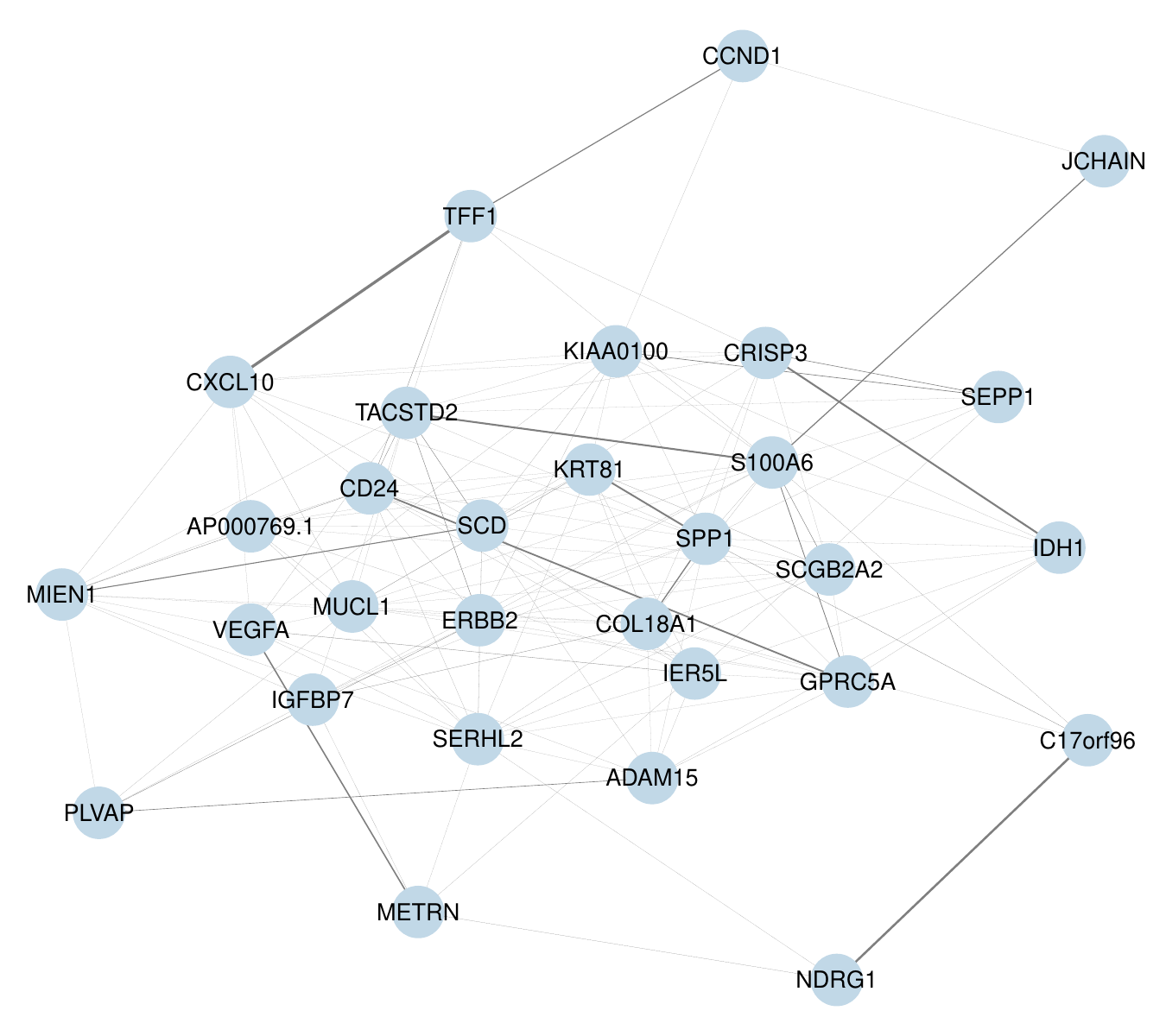}
\caption{Network plot of gene-gene interactions among the top 30 genes selected related to cancer. 
The line between two genes indicates that these genes are related to one another. The thickness of the links represents the magnitude of the interactions.}
\label{fig:da_inter}
\end{figure}

Motivated by these findings, we further investigate whether gene-gene interactions are essential in the cancer development process. We refit model (\ref{eqn:general_equation_ms}) by adding all the two-way interactions among the selected 30 genes to investigate their potential effects on cancer. Then, we select only the interaction terms with effect to standardized deviation ratio greater than 1.96 and display them in Figure \ref{fig:da_inter}, which demonstrates that numerous gene-gene interactions are strongly indicative to the disease status. Notably, we identify several hub genes (e.g. SCD and ERBB2) according to their extensive correlations with many other genes. In addition, VEGFA is known to be related to the proliferation and migration of vascular endothelial cells, which secure a blood supply for growth and are hence upregulated in tumor tissues \citep{qin2023relationship}. Through this gene-gene interaction network, we identify several genes, such as CXCL10 ($\hat{\beta}=-1.0908$, 95\% CrI = $[-1.0948,-1.0867]$) and MIEN1 ($\hat{\beta}=-0.0326$, 95\% CrI = $[-0.0367,-0.0285]$), that potentially collaborate with VEGFA in cancer progression.

\begin{figure}[h]
    \centering
     \begin{subfigure}{0.35\textwidth}
        \centering
        \includegraphics[width=\linewidth]{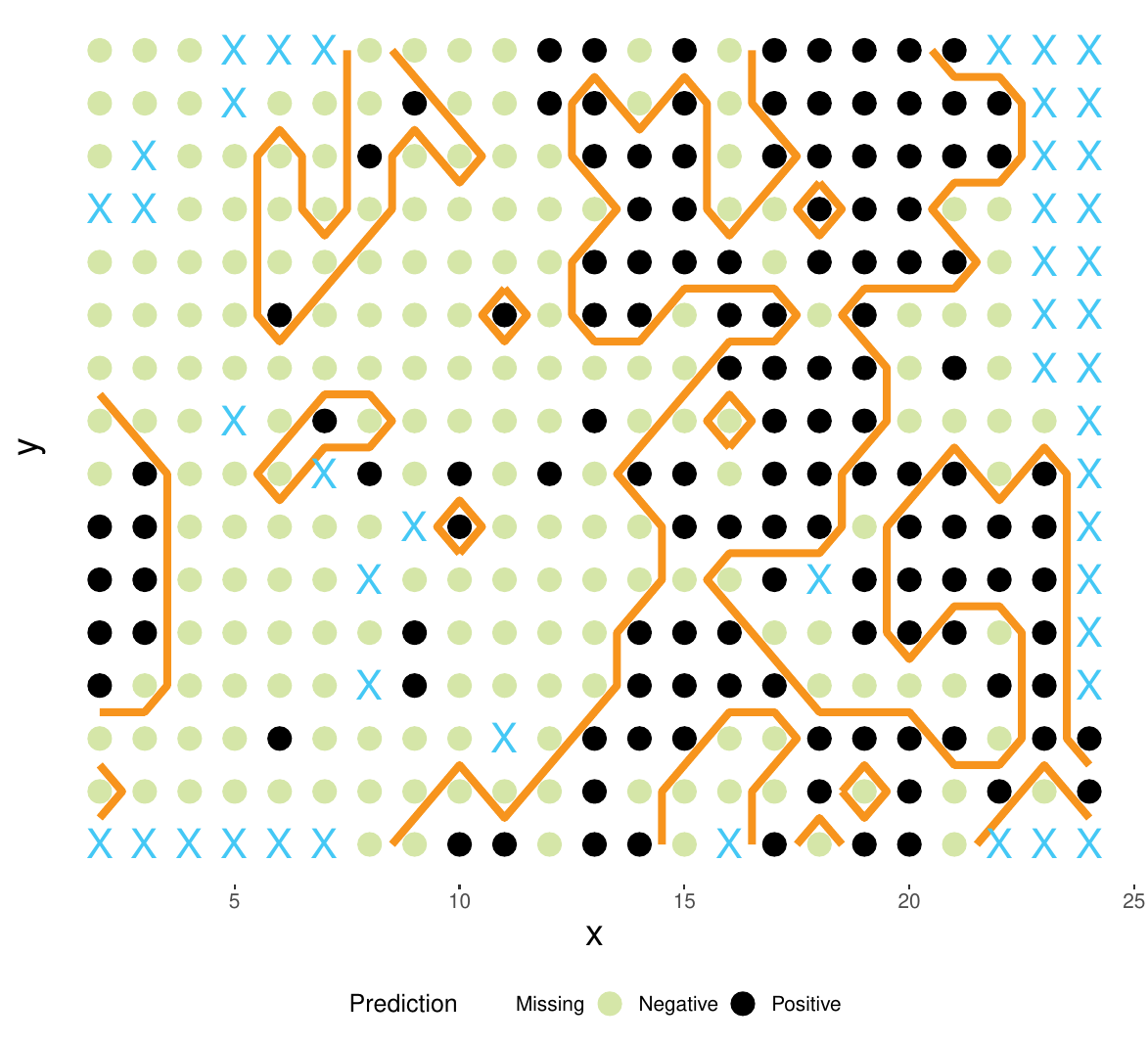}
        \caption{The determinstic prediction by the proposed method.}
        \label{fig:subfig7}
    \end{subfigure}
    \hfill
        \begin{subfigure}{0.35\textwidth}
        \centering
        \includegraphics[width=\linewidth]{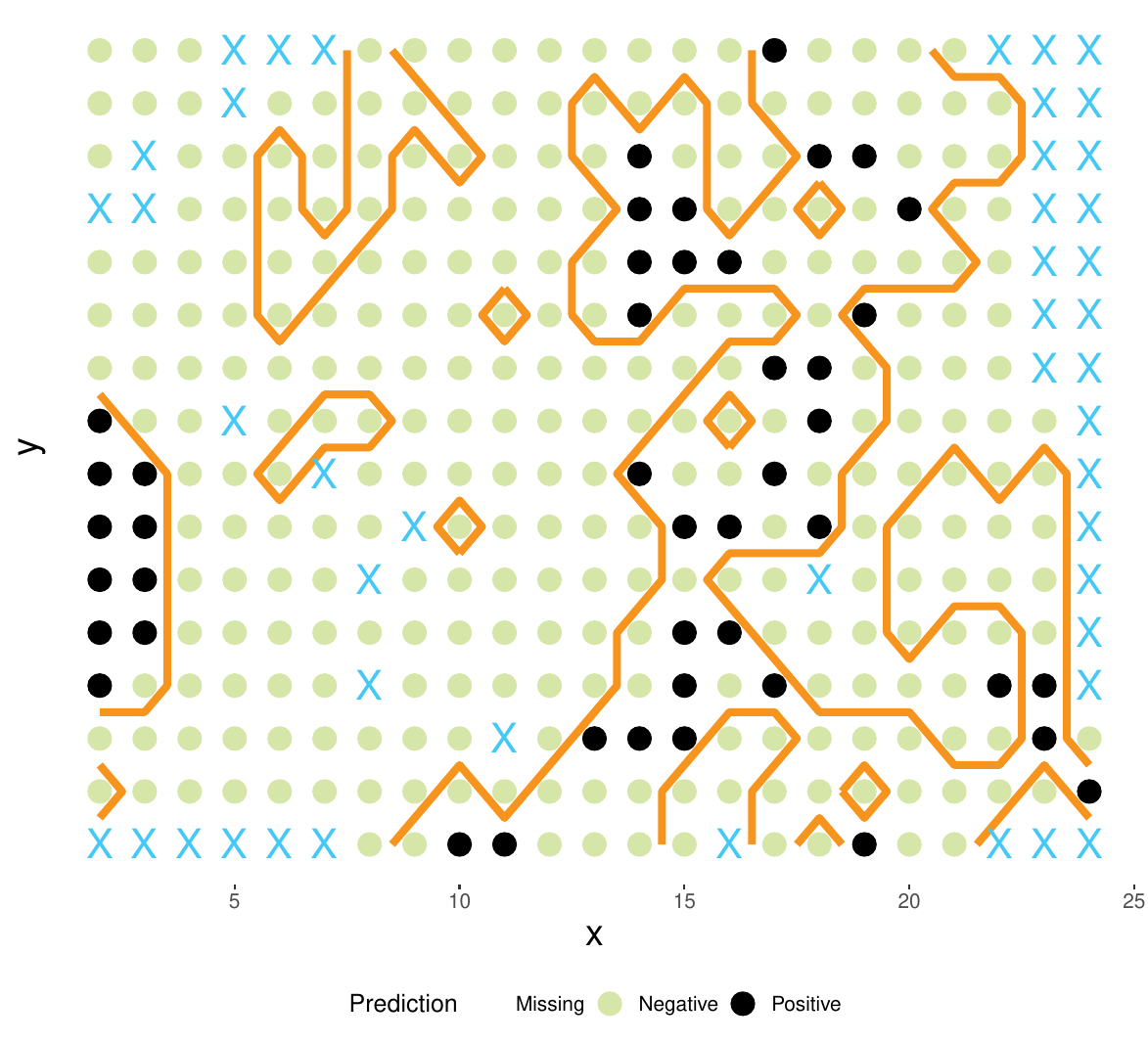}
        \caption{The deterministic prediction by the random forest method.}
        \label{fig:subfig6}
    \end{subfigure}
    \hfill
    \begin{subfigure}{0.35\textwidth}
        \centering
        \includegraphics[width=\linewidth]{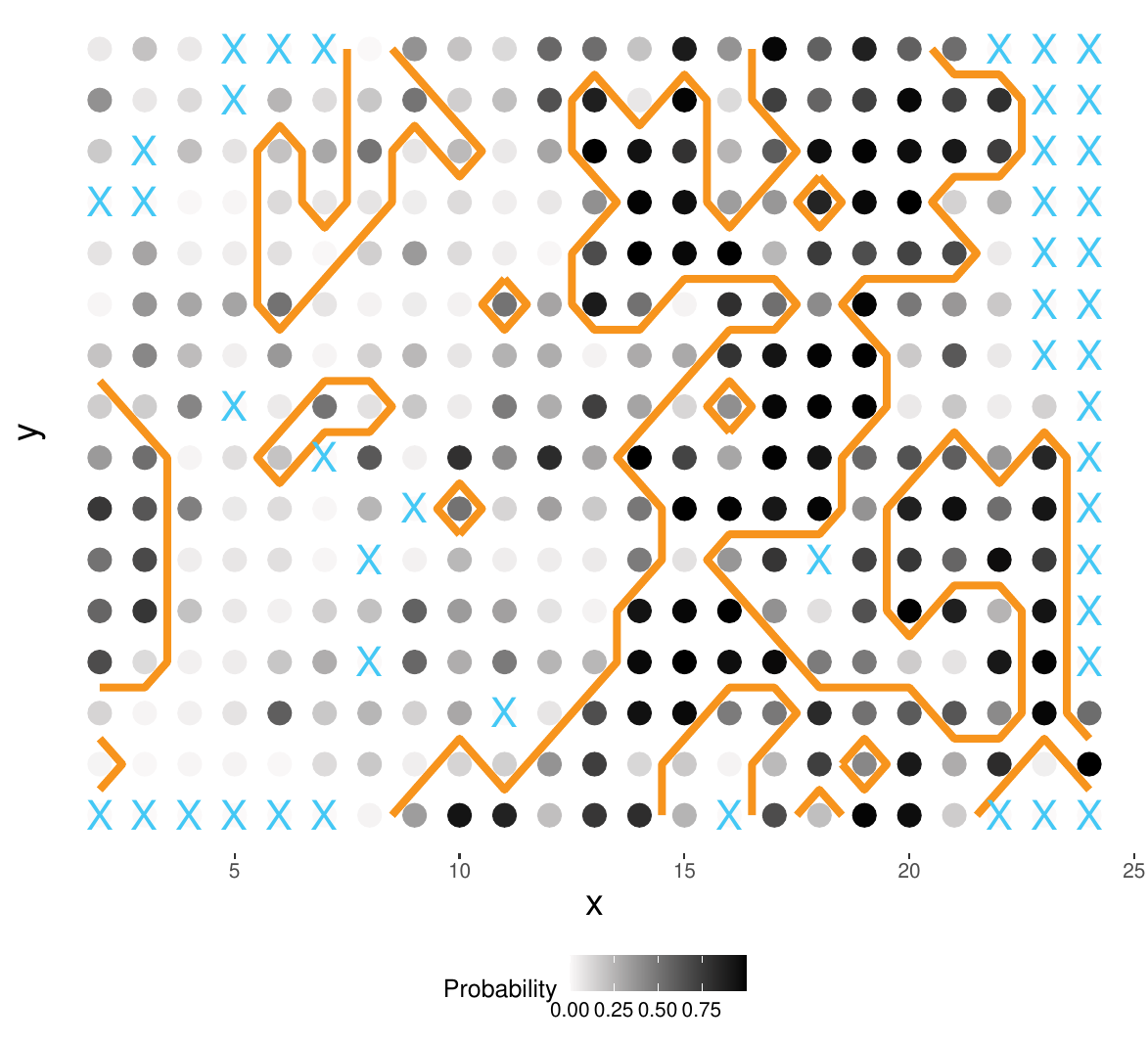}
        \caption{The probabilistic prediction by the proposed method.}
        \label{fig:subfig5}
    \end{subfigure}
    
    \caption{Visualization of predictions by spots on Slice2 of Patient A. (a and b) Deterministic prediction made by the proposed method and the random forest method, where the black and green dots respectively represent the predicted cancerous region and noncancerous region. The orange contour reflects the boundary of the cancerous region and the non-cancerous region. The blue cross (x) represents the missing spots. (c) The probabilistic prediction of $E(Y_i)$. The darkness represents the corresponding predictive probability of being cancerous on that spot.}
    \label{fig:da_countour_pred}
\end{figure}

 To facilitate Objective 2, we further evaluate the predictive accuracy of the proposed model for cancer in an additional tissue sample. We first take Slice 2 of Patient A as the validation data and remove it from the training dataset. Then, we refit the two-way interaction model derived from the thirty genes selected in Objective 1. After fitting the model, predictions are made upon the Slice 2 of Patient A. A spot will be classified into a cancerous region if its $\hat{\mu}$ is greater than a cutoff. The results are presented in Figure \ref{fig:da_countour_pred}. It is evident the prediction of the cancerous region is consistent with the annotation provided by the pathologist. Applying a cutoff of 0.5, the accuracy achieves 0.830. In comparison, we also employ the widely utilized machine-learning-based classification model, Random Forest, for the same predictive procedure. However, the Random Forest only achieves an accuracy of 0.715, as indicated in Figure \ref{fig:subfig6}. This implies that our proposed model not only offers strong interpretability  but also achieves excellent performance in indicating disease status in a given tissue. 
 
In addition to evaluating methods in decision-making as reflected by the deterministic predictions, we also investigate the certainty of predictions across the entire tissue by evaluating the predicted $\hat{\mu}$ at each spot. Typically a prediction is more uncertain when the predicted probability $\hat{\mu}$ is closer to 0.5. Conversely, the prediction is more confident.  As illustrated in Figure \ref{fig:subfig5}, while the prediction is consistent generally with the ground truth annotations, a strong heterogeneity in prediction confidence is observed across the tissue. In particular, spots around the boundary between the cancerous region and the non-cancerous region tend to exhibit a greater uncertainty. This is expected because the transition zone between different tissue types often contains cells with mixed morphological and molecular characteristics, leading to higher ambiguity in classification. Cancerous cells can infiltrate and intermingle with normal cells, creating changes that complicate the distinction between the two regions, thus increasing prediction uncertainty \citep{scimeca2014microcalcifications}.

\section{Discussion}\label{sec:Discussion}

In this article, we introduce a Bayesian shrinkage spatial model that incorporates a spike-and-slab prior to identify the disease-specific genes and construct predictive models to inform cancer diagnosis. The shrinkage prior of the Bayesian spatial model enables the selection of the key relationships by shrinking negligible coefficients to zero. This interpretable approach facilitates both the selection of relevant genes for the disease and the prediction on new data. Additionally, we extend our model to address missing data and model multiple slices simultaneously, thereby enhancing its applicability to various settings of cancer genomics studies and enabling the potential of multi-site analysis. 

We also provide options for users to choose the estimating procedure between NUTS and ADVI, based on their priorities between estimation accuracy and computational expense. This tradeoff is notably important due to the substantial computational expense caused by the large-scale spatial transcriptomics datasets.

This article primarily focuses on sequencing-based spatial transcriptomics where the spots are evenly spaced, and thus the neighbor set of spot $i$ can be straightforwardly defined as the direct adjacent spots of $i$ by letting $\delta=1$. Unlike the spots in sequencing-based methods, the spots in imaging-based spatial transcriptomics are randomly distributed. Nevertheless, the scope of the model is general enough to be adapted for the imaging-based data by carefully specifying $\eta_{ij}$ in Equation (\ref{eqn:general_equation}). For example, one can specify $\delta$ as a larger constant or to be dependent on the spatial distance to the target spot.

Further modifications to our proposed method can be explored to address the misclassifications near the boundary between disease regions, as the misclassification in disease outcomes can often bias the parameter estimation, leading to incorrect conclusions.

\section*{Acknowledgement}
This research was supported by the Natural Sciences and Engineering Research
Council of Canada (NSERC) and Canadian Statistical Sciences Institute (CANSSI) Quebec. Zhang is a Fonds de recherche du Québec Research Scholar (Junior 1). His research was undertaken, in part, thanks to funding from the FRQ-Santé Program.

\newpage
\section*{Appendix}\label{sec:Appendix}

\subsection*{Explicit Form of of $f(\zeta,y|x)$ in Section \ref{subs:trans.cons.var}}

To ease the notion, let $\zeta=T(\theta)=(\Phi^{-1}(\frac{\eta}{c_1}),\beta,log(\tau^2),\Phi^{-1}(w))$, and we have
\begin{equation} \label{eqn:advi_posterior}
    \begin{aligned}
        f(\zeta,y|x)&=\prod_{i=1}^n \mu_i^{y_i} (1-\mu_i)^{1-y_i}\left\{(1-\Phi(\zeta_4) \times \frac{1}{\sqrt{2\pi v_0 e^{\zeta_3}}} \exp\left(-\frac{\beta^2}{2 v_0 e^{\zeta_3}}\right) \times \frac{b_2^{b_1}}{\Gamma(b_1)} e^{\zeta_3(-b_1 - 1)} e^{-\frac{b_2}{x}} \times \frac{1}{c_1}\right.\\
       &\left.+\Phi(\zeta_4) \times \frac{1}{\sqrt{2\pi e^{\zeta_3}}} \exp\left(-\frac{- \mu^2}{2 e^{\zeta_3}}\right) \times \frac{b_2^{b_1}}{\Gamma(b_1)} e^{\zeta_3(-b_1 - 1)} e^{-\frac{b_2}{x}} \times \frac{1}{c_1}\right\} |\text{det} \  J_{T^{-1}}(\zeta)|\\
&=\prod_{i=1}^n \mu_i^{y_i} (1-\mu_i)^{1-y_i}\left\{(1-\Phi(\zeta_4) \times \frac{1}{\sqrt{2\pi v_0 e^{\zeta_3}}} \exp\left(-\frac{\beta^2}{2 v_0 e^{\zeta_3}}\right) \times \frac{b_2^{b_1}}{\Gamma(b_1)} e^{\zeta_3(-b_1 - 1)} e^{-\frac{b_2}{x}} \times \frac{1}{c_1}\right.\\
        &\left.+\Phi(\zeta_4) \times \frac{1}{\sqrt{2\pi e^{\zeta_3}}} \exp\left(-\frac{- \mu^2}{2 e^{\zeta_3}}\right) \times \frac{b_2^{b_1}}{\Gamma(b_1)} e^{\zeta_3(-b_1 - 1)} e^{-\frac{b_2}{x}} \times \frac{1}{c_1}\right\} \times \frac{c_1}{4\pi \sqrt{e^{a^3}}}e^{-\frac{\zeta_1^2+\zeta_4^2}{2}+\zeta^3}
    \end{aligned}
\end{equation}

\noindent where $(\zeta_1,\zeta_2,\zeta_3,\zeta_4)=(\Phi^{-1}(\frac{\eta}{c_1}),\beta,log(\tau^2),\Phi^{-1}(w))$, and $J_{T^{-1}}(\zeta)=\textbf{diag}(\frac{c_1}{\sqrt{2\pi}} e^{-\frac{a_1^2}{2}},1,\frac{e^{a^3}}{2\sqrt{e^{a^3}}},\frac{1}{\sqrt{2\pi}} e^{-\frac{a_4^2}{2}})$ is the Jacobian matrix of the inverse of $T(\theta)$.

\newpage

\newpage

\newpage

\newpage

\newpage

\newpage

\newpage

\newpage
\newpage

\newpage

\newpage

\newpage

\begin{sidewaystable}[!ht]
    \centering
    \captionsetup{font={small,bf,stretch=1.25},justification=raggedright}
    \caption{Results of Simulation for naive method and proposed methods implemented by NUTS and ADVI respectively, with lower correlation $\eta=0.4$}
    \resizebox{1\columnwidth}{!}{\begin{tabular}{c|cccc|cccc|cccc|cccc}
    \hline
        \textbf{} &\textbf{} & \textbf{Naive Method (NUTS)} & \textbf{} & \textbf{} & \textbf{} & \textbf{Proposed Method (NUTS)} & \textbf{} & \textbf{} & \textbf{} & \textbf{Naive Method (ADVI)} & \textbf{} & \textbf{} & \textbf{} & \textbf{Proposed Method(ADVI)} & \textbf{} & \textbf{} \\ \hline
        ~ & avgBias & avgSEE & avgSEM & avgCI &  avgBias & avgSEE & avgSEM & avgCI & avgBias & avgSEE & avgSEM & avgCI &  avgBias & avgSEE & avgSEM & avgCI \\ 
        $\beta_1$ & 0.301 & 0.023 & 0.024 & 94.5\% & 0.089 & 0.016 & 0.016 & 96.0\% & 0.401 & 0.019 & 0.018 & 95.0\% & 0.332 & 0.055 & 0.056 & 95.5\%   \\ 
        $\beta_2$   & 0.648 & 0.024 & 0.025 & 95.0\% & 0.092 & 0.021 & 0.022 & 96.0\% & 0.648 & 0.025 & 0.025 & 95.5\% & 0.353 & 0.047 & 0.047 & 95.5\%  \\ 
        $\beta_3$  & 1.830 & 0.049 & 0.047 & 94.5\% & 0.095 & 0.027 & 0.028 & 95.5\% & 1.830 & 0.038 & 0.038 & 94.5\% & 0.451 & 0.065 & 0.066 & 95.0\% \\ 
        $\beta_4$  & 1.574 & 0.059 & 0.059 & 95.5\% & 0.11 & 0.037 & 0.037 & 95.0\% & 2.274 & 0.059 & 0.058 & 96\% & 0.445 & 0.047 & 0.047 & 95.0\%  \\ 
        $\beta_5$ & 2.196 & 0.063 & 0.064 & 96.0\% & 0.088 & 0.041 & 0.043 & 96.5\%  & 3.296 & 0.064 & 0.066 & 95.0\% & 0.458 & 0.056 & 0.057 & 94.5\% \\ 
        $\beta_6$ & 0.007 & 0.006 & 0.005 & 96.0\% & 0.006 & 0.005 & 0.004 & 96.0\% & 0.139 & 0.008 & 0.008 & 96.0\% & 0.005 & 0.005 & 0.004 & 96.5\%  \\ 
        $\beta_7$ & 0.008 & 0.006 & 0.007 & 95.5\% & 0.007 & 0.005 & 0.005 & 97.5\% & 0.184 & 0.007 & 0.006 & 96.0\% & 0.006 & 0.005 & 0.008 & 94.0\% \\ 
        $\beta_8$ & 0.005 & 0.007 & 0.007 & 96.5\% & 0.006 & 0.005 & 0.005 & 96.0\% & 0.154 & 0.005 & 0.006 & 97.0\% & 0.004 & 0.005 & 0.004 & 95.5\% \\ 
        $\beta_9$ & 0.007 & 0.005 & 0.007 & 97.0\% & 0.006 & 0.006 & 0.005 & 93.0\% & 0.293 & 0.007 & 0.005 & 95.0\% & 0.006 & 0.006 & 0.005 & 95.5\% \\ 
        $\beta_{10}$  & 0.005 & 0.006 & 0.006 & 95.0\% & 0.005 & 0.005 & 0.006 & 95.5\% & 0.231 & 0.006 & 0.005 & 94.0\% & 0.005 & 0.005 & 0.004 & 95.0\% \\ 
        $\beta_{11}$ & 0.006 & 0.008 & 0.010 & 94.5\% & 0.006 & 0.005 & 0.006 & 96.0\%  & 0.142 & 0.006 & 0.06 & 95.5\% & 0.006 & 0.004 & 0.004 & 96.0\% \\ 
        $\beta_{12}$  & 0.004 & 0.007 & 0.008 & 97.0\% & 0.006 & 0.004 & 0.005 & 95.0\%  & 0.185 & 0.008 & 0.007 & 95.0\% & 0.005 & 0.005 & 0.006 & 95.0\% \\ 
        $\beta_{13}$ & 0.005 & 0.005 & 0.006 & 95.0\% & 0.004 & 0.005 & 0.006 & 95.5\% & 0.283 & 0.008 & 0.008 & 98.0\% & 0.004 & 0.006 & 0.006 & 94.0\% \\ 
        $\beta_{14}$  & 0.005 & 0.007 & 0.008 & 95.5\% & 0.007 & 0.006 & 0.007 & 97.0\% & 0.231 & 0.008 & 0.008 & 95.0\% & 0.005 & 0.006 & 0.006 & 93.0\% \\ 
        $\beta_{15}$ & 0.006 & 0.008 & 0.006 & 97.0\% & 0.007 & 0.005 & 0.005 & 95.0\%  & 0.125 & 0.008 & 0.008 & 95.5\% & 0.005 & 0.004 & 0.005 & 94.5\%  \\ 
        $\beta_{16}$ & 0.006 & 0.005 & 0.007 & 96.5\% & 0.006 & 0.006 & 0.006 & 95.0\%  & 0.193 & 0.006 & 0.006 & 96.0\% & 0.005 & 0.006 & 0.009 & 96.5\% \\ 
        $\beta_{17}$ & 0.004 & 0.006 & 0.006 & 96.0\% & 0.007 & 0.005 & 0.007 & 94.0\%  & 0.205 & 0.007 & 0.008 & 96.0\% & 0.005 & 0.005 & 0.006 & 94.5\% \\ 
        $\beta_{18}$ & 0.005 & 0.007 & 0.009 & 94.5\% & 0.006 & 0.006 & 0.005 & 95.0\%  & 0.153 & 0.007 & 0.08 & 96.5\% & 0.006 & 0.006 & 0.004 & 95.0\%  \\ 
        $\beta_{19}$  & 0.006 & 0.008 & 0.011 & 95.5\% & 0.005 & 0.006 & 0.006 & 94.0\% & 0.139 & 0.006 & 0.008 & 96.5\% & 0.005 & 0.006 & 0.007 & 95.0\%\\ 
        $\beta_{20}$ & 0.007 & 0.007 & 0.008 & 98.0\% & 0.006 & 0.006 & 0.007 & 97.0\% & 0.205 & 0.006 & 0.009 & 95.5\% & 0.005 & 0.005 & 0.006 & 94.5\%  \\ 
        $\eta$  & $-$ & $-$ & $-$ & $-$ & 0.019 & 0.021 & 0.022 & 95.0\%  & $-$ & $-$ & $-$ & $-$ & 0.087 & 0.044 & 0.044 & 95.0\% \\ \hline 
    \end{tabular}}
    \label{tab:sim1_1}
\end{sidewaystable}
\begin{sidewaystable}[!ht]
    \centering
    \captionsetup{font={small,bf,stretch=1.25},justification=raggedright}
    \caption{Results of Simulation for naive method and proposed methods implemented by NUTS and ADVI respectively, with lower correlation $\eta=1.6$}
    \resizebox{1\columnwidth}{!}{\begin{tabular}{c|cccc|cccc|cccc|cccc}
    \hline
        \textbf{} & \textbf{} & \textbf{Naive Method (NUTS)} & \textbf{} & \textbf{} & \textbf{} & \textbf{Proposed Method (NUTS)} & \textbf{} & \textbf{} & \textbf{} & \textbf{Naive Method (ADVI)} & \textbf{} & \textbf{} & \textbf{} & \textbf{Proposed Method(ADVI)} & \textbf{} & \textbf{} \\ \hline
        ~ & avgBias & avgSEE & avgSEM & avgCI &  avgBias & avgSEE & avgSEM & avgCI & avgBias & avgSEE & avgSEM & avgCI &  avgBias & avgSEE & avgSEM & avgCI \\ 
        $\beta_1$ & 0.343 & 0.016 & 0.018 & 94.0\% & 0.083 & 0.011 & 0.013 & 95.5\% & 0.389 & 0.016 & 0.016 & 95.5\% & 0.353 & 0.043 & 0.043 & 93.5\%   \\ 
        $\beta_2$    & 0.447 & 0.023 & 0.027 & 95.0\% & 0.073 & 0.017 & 0.017 & 95.0\%  & 0.689 & 0.025 & 0.027 & 95.0\% & 0.443 & 0.032 & 0.033 & 95.5\%  \\ 
        $\beta_3$  & 1.192 & 0.034 & 0.034 & 95.0\% & 0.095 & 0.015 & 0.029 & 96.0\%  & 1.540 & 0.034 & 0.034 & 94.0\% & 0.355 & 0.033 & 0.034 & 94.5\% \\ 
        $\beta_4$  & 1.738 & 0.067 & 0.065 & 95.5\% & 0.082 & 0.033 & 0.034 & 95.0\% & 2.386 & 0.067 & 0.067 & 95.0\% & 0.452 & 0.046 & 0.047 & 96.0\%  \\ 
        $\beta_5$ & 1.823 & 0.069 & 0.069 & 95.0\% & 0.183 & 0.04 & 0.043 & 94.5\%   & 3.982 & 0.069 & 0.069 & 95.5\% & 0.447 & 0.038 & 0.039 & 96.5\% \\ 
        $\beta_6$ & 0.005 & 0.004 & 0.004 & 97.0\% & 0.006 & 0.004 & 0.004 & 95.0\% & 0.138 & 0.007 & 0.007 & 95.0\% & 0.006 & 0.008 & 0.008 & 95.0\% \\ 
        $\beta_7$ & 0.008 & 0.006 & 0.007 & 95.0\% & 0.004 & 0.004 & 0.008 & 97.0\%  & 0.174 & 0.008 & 0.008 & 94.5\% & 0.005 & 0.009 & 0.008 & 94.5\% \\ 
        $\beta_8$ & 0.005 & 0.009 & 0.009 & 96.5\% & 0.005 & 0.005 & 0.004 & 95.5\%  & 0.164 & 0.006 & 0.007 & 96.0\% & 0.008 & 0.009 & 0.009 & 96.5\% \\ 
        $\beta_9$ & 0.006 & 0.007 & 0.009 & 96.5\% & 0.004 & 0.004 & 0.005 & 96.5\% & 0.193 & 0.006 & 0.007 & 97.0\% & 0.007 & 0.008 & 0.009 & 95.0\% \\ 
        $\beta_{10}$  & 0.005 & 0.007 & 0.007 & 93.0\% & 0.006 & 0.004 & 0.004 & 96.0\% & 0.271 & 0.006 & 0.006 & 96.0\% & 0.007 & 0.007 & 0.009 & 95.5\%\\ 
        $\beta_{11}$ & 0.005 & 0.007 & 0.011 & 97.5\% & 0.006 & 0.004 & 0.004 & 98.0\%  & 0.192 & 0.008 & 0.08 & 95.0\% & 0.008 & 0.008 & 0.009 & 95.5\% \\ 
        $\beta_{12}$  & 0.007 & 0.006 & 0.007 & 96.0\% & 0.007 & 0.004 & 0.006 & 95.5\% & 0.185 & 0.006 & 0.007 & 95.5\% & 0.007 & 0.006 & 0.006 & 95.5\%  \\ 
        $\beta_{13}$ & 0.007 & 0.008 & 0.007 & 96.0\% & 0.005 & 0.006 & 0.006 & 96.0\% & 0.283 & 0.006 & 0.006 & 96.0\% & 0.007 & 0.007 & 0.005 & 94.5\% \\ 
        $\beta_{14}$   & 0.006 & 0.008 & 0.008 & 96.0\% & 0.005 & 0.004 & 0.006 & 95.0\% & 0.131 & 0.007 & 0.007 & 95.5\% & 0.007 & 0.005 & 0.005 & 96.0\% \\ 
        $\beta_{15}$  & 0.007 & 0.007 & 0.008 & 94.5\% & 0.006 & 0.005 & 0.005 & 95.5\%  & 0.222 & 0.007 & 0.006 & 95.5\% & 0.006 & 0.009 & 0.007 & 95.0\% \\ 
        $\beta_{16}$ & 0.007 & 0.007 & 0.008 & 94.0\% & 0.005 & 0.005 & 0.009 & 95.0\%  & 0.187 & 0.007 & 0.007 & 97.0\% & 0.007 & 0.009 & 0.008 & 94.5\% \\ 
        $\beta_{17}$ & 0.008 & 0.006 & 0.008 & 97.0\% & 0.005 & 0.004 & 0.006 & 94.0\%  & 0.195 & 0.006 & 0.006 & 94.5\% & 0.007 & 0.009 & 0.008 & 95.0\%\\ 
        $\beta_{18}$  & 0.005 & 0.007 & 0.013 & 94.0\% & 0.004 & 0.004 & 0.004 & 96.5\%  & 0.183 & 0.008 & 0.008 & 94.0\% & 0.006 & 0.009 & 0.009 & 95.0\%   \\ 
        $\beta_{19}$   & 0.006 & 0.007 & 0.008 & 95.5\% & 0.005 & 0.005 & 0.007 & 94.0\% & 0.169 & 0.006 & 0.006 & 95.5\% & 0.008 & 0.008 & 0.008 & 95.0\% \\ 
        $\beta_{20}$ & 0.006 & 0.007 & 0.009 & 98.0\% & 0.005 & 0.005 & 0.006 & 95.5\% & 0.198 & 0.008 & 0.007 & 95.0\% & 0.008 & 0.009 & 0.009 & 95.5\%  \\ 
        $\eta$  & $-$ & $-$ & $-$ & $-$ & 0.021 & 0.028 & 0.030 & 96.0\%    & $-$ & $-$ & $-$ & $-$ & 0.083 & 0.069 & 0.070 & 95.0\% \\ \hline 
    \end{tabular}}
    \label{tab:sim1_2}
\end{sidewaystable}
\begin{sidewaystable}[!ht]
    \centering
    \captionsetup{font={small,bf,stretch=1.25},justification=raggedright}
    \caption{Results of Simulation for naive method and proposed methods implemented by NUTS and ADVI respectively, with lower correlation $\eta=2.8$}
    \resizebox{1\columnwidth}{!}{\begin{tabular}{c|cccc|cccc|cccc|cccc}
    \hline
        \textbf{} & \textbf{} & \textbf{Naive Method (NUTS)} & \textbf{} & \textbf{} & \textbf{} & \textbf{Proposed Method (NUTS)} & \textbf{} & \textbf{} & \textbf{} & \textbf{Naive Method (ADVI)} & \textbf{} & \textbf{} & \textbf{} & \textbf{Proposed Method(ADVI)} & \textbf{} & \textbf{} \\ \hline
        ~ & avgBias & avgSEE & avgSEM & avgCI &  avgBias & avgSEE & avgSEM & avgCI & avgBias & avgSEE & avgSEM & avgCI &  avgBias & avgSEE & avgSEM & avgCI \\ 
        $\beta_1$ & 0.037 & 0.038 & 0.038 & 95.0\% & 0.034 & 0.056 & 0.057 & 96.5\% & 0.423 & 0.017 & 0.017 & 94.0\% & 0.334 & 0.055 & 0.056 & 94.5\%   \\ 
        $\beta_2$    & 0.043 & 0.045 & 0.043 & 97.0\% & 0.067 & 0.062 & 0.063 & 95.0\%  & 0.582 & 0.027 & 0.027 & 94.5\% & 0.467 & 0.047 & 0.047 & 96.5\%  \\ 
        $\beta_3$   & 1.273 & 0.036 & 0.032 & 95.5\% & 0.044 & 0.058 & 0.058 & 95.0\%  & 1.673 & 0.038 & 0.037 & 95.5\% & 0.344 & 0.065 & 0.066 & 95.5\% \\ 
        $\beta_4$   & 1.833 & 0.062 & 0.062 & 94.5\% & 0.056 & 0.037 & 0.037 & 93.0\% & 2.833 & 0.063 & 0.063 & 97.0\% & 0.356 & 0.047 & 0.047 & 94.0\%  \\ 
        $\beta_5$ & 1.786 & 0.073 & 0.072 & 95.0\% & 0.056 & 0.058 & 0.059 & 94.5\%   & 3.786 & 0.072 & 0.073 & 95.5\% & 0.556 & 0.056 & 0.057 & 94.0\% \\ 
        $\beta_6$ & 0.008 & 0.006 & 0.007 & 96.0\% & 0.008 & 0.006 & 0.005 & 95.5\% & 0.133 & 0.006 & 0.006 & 94.0\% & 0.008 & 0.007 & 0.007 & 96.0\% \\ 
        $\beta_7$ & 0.005 & 0.005 & 0.004 & 96.0\% & 0.005 & 0.005 & 0.006 & 96.0\%  & 0.144 & 0.007 & 0.007 & 95.5\% & 0.008 & 0.006 & 0.006 & 94.5\% \\ 
        $\beta_8$ & 0.008 & 0.006 & 0.005 & 94.0\% & 0.008 & 0.007 & 0.007 & 94.0\%   & 0.184 & 0.006 & 0.007 & 95.0\% & 0.007 & 0.007 & 0.007 & 95.0\% \\ 
        $\beta_9$ & 0.006 & 0.006 & 0.006 & 94.5\% & 0.0008 & 0.005 & 0.006 & 96.0\% & 0.195 & 0.005 & 0.006 & 95.5\% & 0.004 & 0.007 & 0.005 & 95.0\% \\ 
        $\beta_{10}$  & 0.006 & 0.007 & 0.006 & 95.5\% & 0.007 & 0.007 & 0.006 & 96.5\% & 0.168 & 0.007 & 0.006 & 96.0\% & 0.005 & 0.007 & 0.007 & 95.5\% \\ 
        $\beta_{11}$ & 0.007 & 0.005 & 0.008 & 94.0\% & 0.006 & 0.004 & 0.006 & 95.0\%  & 0.242 & 0.006 & 0.07 & 95.0\% & 0.005 & 0.006 & 0.006 & 94.5\% \\ 
        $\beta_{12}$ & 0.005 & 0.006 & 0.006 & 97.5\% & 0.008 & 0.006 & 0.007 & 95.0\% & 0.172 & 0.005 & 0.008 & 94.5\% & 0.007 & 0.007 & 0.008 & 95.5\%  \\ 
        $\beta_{13}$ & 0.006 & 0.006 & 0.007 & 95.5\% & 0.009 & 0.007 & 0.006 & 97.0\%  & 0.194 & 0.008 & 0.008 & 95.5\% & 0.008 & 0.008 & 0.008 & 93.0\% \\ 
        $\beta_{14}$   & 0.006 & 0.006 & 0.005 & 95.5\% & 0.007 & 0.008 & 0.007 & 95.5\% & 0.287 & 0.009 & 0.009 & 95.0\% & 0.007 & 0.005 & 0.005 & 93.5\% \\ 
        $\beta_{15}$ & 0.005 & 0.005 & 0.006 & 97.5\% & 0.008 & 0.008 & 0.008 & 95.0\%  & 0.276 & 0.007 & 0.008 & 93.5\% & 0.005 & 0.007 & 0.008 & 95.5\% \\ 
        $\beta_{16}$ & 0.006 & 0.005 & 0.007 & 94.0\% & 0.007 & 0.009 & 0.009 & 94.0\%  & 0.217 & 0.008 & 0.008 & 95.0\% & 0.007 & 0.007 & 0.005 & 95.5\% \\ 
        $\beta_{17}$ & 0.006 & 0.005 & 0.006 & 95.0\% & 0.008 & 0.006 & 0.007 & 94.5\%   & 0.198 & 0.007 & 0.007 & 94.5\% & 0.008 & 0.006 & 0.006 & 95.5\%\\ 
        $\beta_{18}$  & 0.006 & 0.007 & 0.007 & 95.5\% & 0.008 & 0.007 & 0.005 & 96.5\%  & 0.249 & 0.007 & 0.006 & 93.5\% & 0.007 & 0.008 & 0.008 & 95.5\%   \\ 
        $\beta_{19}$   & 0.004 & 0.006 & 0.009 & 95.0\% & 0.008 & 0.005 & 0.006 & 95.0\% & 0.219 & 0.006 & 0.006 & 96.0\% & 0.007 & 0.007 & 0.007 & 96.0\% \\ 
        $\beta_{20}$ & 0.005 & 0.007 & 0.007 & 96.0\% & 0.006 & 0.007 & 0.007 & 94.5\%  & 0.226 & 0.007 & 0.008 & 94.5\% & 0.007 & 0.007 & 0.007 & 96.5\%  \\ 
        $\eta$  & $-$ & $-$ & $-$ & $-$ & 0.045 & 0.047 & 0.045 & 95.5\%    & $-$ & $-$ & $-$ & $-$ & 0.095 & 0.057 & 0.056 & 95.5\% \\ \hline 
    \end{tabular}}
    \label{tab:sim1_3}
\end{sidewaystable}

\newpage

\begin{sidewaystable}[!ht]
    \centering
    \captionsetup{font={small,bf,stretch=1.25},justification=raggedright}
    \caption{Simulation result}
    \resizebox{1\columnwidth}{!}{\begin{tabular}{c|cccc|cccc|cccc|cccc}
    \hline
        \textbf{} & \textbf{} & \textbf{$\sigma=0.1, \rho=0.1$} & \textbf{} & \textbf{} & \textbf{} & \textbf{$\sigma=0.1,\rho=0.4$} & \textbf{} & \textbf{} & \textbf{} & \textbf{$\sigma=0.4,\rho=0.1$} & \textbf{} & \textbf{} & \textbf{} & \textbf{$\sigma=0.4,\rho=0.4$} & \textbf{} & \textbf{} \\ \hline
        ~ & avgBias & avgSEE & avgSEM & avgCI &  avgBias & avgSEE & avgSEM & avgCI & avgBias & avgSEE & avgSEM & avgCI &  avgBias & avgSEE & avgSEM & avgCI \\ 
        $\beta_1$ & 0.303 & 0.046 & 0.047 & 95.5\% & 0.362 & 0.045 & 0.043 & 94.0\% & 0.288 & 0.034 & 0.034 & 95.0\% & 0.371 & 0.032 & 0.033 & 94.5\%   \\ 
        $\beta_2$  & 0.331 & 0.032 & 0.032 & 94.0\% & 0.481 & 0.054 & 0.054 & 95.0\% & 0.351 & 0.045 & 0.045 & 94.0\% & 0.323 & 0.059 & 0.059 & 95.5\% \\ 
        $\beta_3$ & 0.398 & 0.048 & 0.048 & 95.5\% & 0.452 & 0.042 & 0.043 & 94.5\% & 0.424 & 0.046 & 0.045 & 94.0\% & 0.511 & 0.048 & 0.047 & 93.0\% \\ 
        $\beta_4$ & 0.494 & 0.057 & 0.059 & 96.5\% & 0.568 & 0.076 & 0.074 & 95.5\% & 0.412 & 0.057 & 0.054 & 95.0\% & 0.493 & 0.049 & 0.045 & 95.0\%  \\ 
        $\beta_5$ & 0.469 & 0.043 & 0.043 & 95.0\% & 0.524 & 0.054 & 0.054 & 93.5\% & 0.459 & 0.063 & 0.063 & 95.5\% & 0.551 & 0.063 & 0.059 & 94.5\% \\ 
        $\beta_6$ & 0.008 & 0.005 & 0.005 & 94.0\% & 0.005 & 0.009 & 0.008 & 93.0\% & 0.006 & 0.008 & 0.008 & 95.0\% & 0.004 & 0.009 & 0.009 & 94.0\%  \\ 
        $\beta_7$ & 0.007 & 0.006 & 0.006 & 94.5\% & 0.008 & 0.009 & 0.009 & 92.5\% & 0.007 & 0.009 & 0.007 & 93.0\% & 0.006 & 0.009 & 0.009 & 92.0\% \\ 
        $\beta_8$ & 0.008 & 0.009 & 0.009 & 97.0\% & 0.007 & 0.008 & 0.008 & 95.5\%  & 0.009 & 0.009 & 0.009 & 94.5\% & 0.007 & 0.010 & 0.010 & 95.5\% \\ 
        $\beta_9$ & 0.008 & 0.008 & 0.008 & 96.0\% & 0.007 & 0.009 & 0.009 & 94.0\% & 0.006& 0.009 & 0.009 & 96.0\% & 0.095 & 0.009 & 0.009 & 94.5\% \\ 
        $\beta_{10}$ & 0.006 & 0.008 & 0.008 & 93.0\% & 0.007 & 0.009 & 0.009 & 96.0\% & 0.009 & 0.007 & 0.007 & 95.5\% & 0.008 & 0.011 & 0.011 & 94.0\% \\ 
        $\beta_{11}$ & 0.007 & 0.009 & 0.008 & 95.5\% & 0.008 & 0.009 & 0.009 & 95.0\% & 0.008 & 0.009 & 0.010 & 93.5\% & 0.009 & 0.008 & 0.008 & 92.0\% \\ 
        $\beta_{12}$  & 0.007 & 0.009 & 0.007 & 95.0\% & 0.008 & 0.009 & 0.007 & 96.5\% & 0.005 & 0.008 & 0.008 & 94.0\% & 0.009 & 0.008 & 0.008 & 95.5\% \\ 
        $\beta_{13}$ & 0.008 & 0.006 & 0.006 & 93.0\% & 0.009 & 0.009 & 0.009 & 94.5\% & 0.009 & 0.008 & 0.007 & 92.0\% & 0.007 & 0.006 & 0.006 & 93.0\% \\ 
        $\beta_{14}$ & 0.008 & 0.008 & 0.008 & 93.5\% & 0.009 & 0.008 & 0.009 & 93.0\% & 0.009 & 0.008 & 0.008 & 93.0\% & 0.007 & 0.006 & 0.006 & 94.5\% \\ 
        $\beta_{15}$ & 0.009 & 0.008 & 0.008 & 94.0\% & 0.011 & 0.010 & 0.011 & 95.5\% & 0.008 & 0.009 & 0.009 & 95.5\% & 0.012 & 0.009 & 0.009 & 96.5\%  \\ 
        $\beta_{16}$ & 0.009 & 0.008 & 0.007 & 95.5\% & 0.09 & 0.009 & 0.009 & 93.5\% & 0.007& 0.009 & 0.009 & 96.0\% & 0.007 & 0.009 & 0.009 & 93.0\% \\ 
        $\beta_{17}$ & 0.009 & 0.008 & 0.006 & 94.0\% & 0.009 & 0.009 & 0.009 & 95.0\% & 0.007 & 0.008 & 0.008 & 94.0\% & 0.008 & 0.009 & 0.009 & 95.0\% \\ 
        $\beta_{18}$ & 0.008 & 0.006 & 0.006 & 94.0\% & 0.007 & 0.009 & 0.009 & 95.5\% & 0.004 & 0.008 & 0.011& 96.0\% & 0.009 & 0.009 & 0.011 & 93.5\%  \\ 
        $\beta_{19}$ & 0.007 & 0.008 & 0.005 & 93.5\% & 0.008 & 0.009 & 0.009 & 95.0\% & 0.004 & 0.008 & 0.008 & 94.5\% & 0.009 & 0.009 & 0.007 & 96.0\% \\ 
        $\beta_{20}$ & 0.005 & 0.009 & 0.008 & 95.0\% & 0.009 & 0.007 & 0.007 & 95.0\% & 0.009 & 0.008 & 0.008 & 94.0\% & 0.012 & 0.009 & 0.007 & 91.5\%  \\ 
        $\eta$ & 0.076 & 0.025 & 0.025 & 94.5\% & 0.089 & 0.031 & 0.031 & 94.0\%  & 0.083 & 0.022 & 0.022 & 93.5\% & 0.081 & 0.029 & 0.026 & 95.0\% \\ \hline 
    \end{tabular}}
    \label{tab:sim3_bias}
\end{sidewaystable}

\newpage

\begin{table}[!ht]
    \centering
    \captionsetup{font={small,bf,stretch=1.25},justification=raggedright}
    \caption{Result for Simulation in the ignorable missing setting}
    \resizebox{.9\columnwidth}{!}{\begin{tabular}{c|cccc|cccc }
    \hline
        \textbf{} & \textbf{} & \textbf{10 Missing Spots} & \textbf{} & \textbf{} & \textbf{} & \textbf{30 Missing Spots} & \textbf{} & \textbf{} \\ \hline
        ~ & avgBias & avgSEE & avgSEM & avgCI &  avgBias & avgSEE & avgSEM & avgCI \\ 
        $\beta_1$ & 0.335 & 0.043 & 0.043 & 95.0\% & 0.535 & 0.028 & 0.028 & 97.0\% \\ 
        $\beta_2$ & 0.449 & 0.056 & 0.057 & 95.0\% & 0.796 & 0.028 & 0.029 & 95.5\% \\ 
        $\beta_3$ & 0.465 & 0.048 & 0.048 & 95.5\% & 0.664 & 0.038 & 0.038 & 95.0\% \\ 
        $\beta_4$ & 0.536 & 0.039 & 0.039 & 96.5\% & 0.816 & 0.048 & 0.047 & 96.5\% \\ 
        $\beta_5$ & 0.552 & 0.055 & 0.056 & 93.5\% & 0.787 & 0.047 & 0.047 & 95.5\% \\ 
        $\beta_6$ & 0.009 & 0.006 & 0.006 & 95.0\% & 0.009 & 0.009 & 0.009 & 94.5\% \\ 
        $\beta_7$ & 0.009 & 0.008 & 0.007 & 96.5\% & 0.009 & 0.009 & 0.009 & 96.0\% \\ 
        $\beta_8$ & 0.008 & 0.006 & 0.006 & 94.0\%  & 0.011 & 0.009 & 0.009 & 94.5\% \\ 
        $\beta_9$ & 0.009 & 0.006 & 0.005 & 95.0\% & 0.009 & 0.007 & 0.009 & 96.0\%  \\ 
        $\beta_{10}$ & 0.008 & 0.007 & 0.007 & 96.0\%  & 0.009 & 0.008 & 0.008 & 97.0\% \\ 
        $\beta_{11}$ & 0.007 & 0.008 & 0.008 & 96.5\% & 0.008 & 0.009 & 0.009 & 96.0\% \\ 
        $\beta_{12}$ & 0.009 & 0.008 & 0.008 & 93.5\% & 0.009 & 0.007 & 0.007 & 95.0\% \\ 
        $\beta_{13}$  & 0.010 & 0.009 & 0.009 & 96.0\% & 0.012 & 0.008 & 0.007 & 95.0\% \\ 
        $\beta_{14}$ & 0.008 & 0.009 & 0.009 & 97.0\% & 0.009 & 0.006 & 0.006 & 96.0\% \\ 
        $\beta_{15}$ & 0.006 & 0.009 & 0.010 & 94.5\%  & 0.009 & 0.008 & 0.009 & 95.0\% \\ 
        $\beta_{16}$ & 0.008 & 0.009 & 0.009 & 97.0\% & 0.007 & 0.007 & 0.007 & 95.0\% \\ 
        $\beta_{17}$ & 0.007 & 0.009 & 0.010 & 94.0\% & 0.008 & 0.008 & 0.008 & 95.5\% \\ 
        $\beta_{18}$  & 0.008 & 0.004 & 0.007 & 97.5\% & 0.009 & 0.009 & 0.009 & 97.5\% \\ 
        $\beta_{19}$ & 0.008 & 0.008 & 0.009 & 95.0\% & 0.008 & 0.006 & 0.007 & 95.5\% \\ 
        $\beta_{20}$ & 0.008 & 0.009 & 0.008 & 95.5\% & 0.009 & 0.007 & 0.008 & 95.5\% \\ 
        $\eta$ &  0.089 & 0.011 & 0.012 & 95.5\% & 0.156 & 0.029 & 0.029 & 95.0\% \\ \hline
    \end{tabular}}
    \label{sim3_1}
\end{table}
\begin{table}[!ht]
    \centering
    \captionsetup{font={small,bf,stretch=1.25},justification=raggedright}
    \caption{Result for Simulation in the nonignorable missing setting}
    \resizebox{.9\columnwidth}{!}{\begin{tabular}{c|cccc|cccc }
    \hline
        \textbf{} & \textbf{} & \textbf{$(\gamma_0,\gamma_1,\gamma_2)^{\intercal}=(-6,1,4)^{\intercal}$} & \textbf{} & \textbf{} & \textbf{} & \textbf{$(\gamma_0,\gamma_1,\gamma_2)^{\intercal}=(-5,1,1.6)^{\intercal}$} & \textbf{} & \textbf{} \\ \hline
        ~ & avgBias & avgSEE & avgSEM & avgCI &  avgBias & avgSEE & avgSEM & avgCI \\ 
        $\beta_1$ & 0.398 & 0.041 & 0.041 & 94.5\%  & 0.438 & 0.046 & 0.045 & 95.0\% \\ 
        $\beta_2$ & 0.349 & 0.037 & 0.037 & 94.0\% & 0.451 & 0.048 & 0.047 & 95.5\% \\ 
        $\beta_3$ & 0.489 & 0.048 & 0.048 & 94.0\% & 0.511 & 0.053 & 0.053 & 96.0\% \\ 
        $\beta_4$ & 0.501 & 0.050 & 0.051 & 95.5\% & 0.519 & 0.042& 0.045 & 95.5\% \\ 
        $\beta_5$ & 0.471 & 0.059 & 0.056 & 95.5\% & 0.507 & 0.055 & 0.055 & 92.5\% \\ 
        $\beta_6$ & 0.009 & 0.009 & 0.009 & 96.5\% & 0.008 & 0.008 & 0.008 & 91.5\% \\ 
        $\beta_7$ & 0.009 & 0.009 & 0.009 & 97.5\% & 0.008 & 0.007 & 0.007 & 93.5\% \\ 
        $\beta_8$ & 0.008 & 0.008 & 0.008 & 95.5\%  & 0.009 & 0.009 & 0.009 & 94.5\% \\ 
        $\beta_9$ & 0.009 & 0.008 & 0.008 & 95.5\% & 0.008 & 0.006 & 0.006 & 95.0\%  \\ 
        $\beta_{10}$ & 0.009 & 0.008 & 0.007 & 93.0\%  & 0.009 & 0.006 & 0.005 & 94.0\% \\ 
        $\beta_{11}$ & 0.006 & 0.006 & 0.006 & 94.5\% & 0.008 & 0.007 & 0.006 & 95.5\% \\ 
        $\beta_{12}$ & 0.007 & 0.005 & 0.005 & 97.0\% & 0.008 & 0.005 & 0.006 & 95.0\% \\ 
        $\beta_{13}$  & 0.007 & 0.006 & 0.009 & 95.0\% & 0.009 & 0.007 & 0.009 & 94.0\% \\ 
        $\beta_{14}$ & 0.005 & 0.008 & 0.008 & 95.0\% & 0.009 & 0.009 & 0.009 & 93.0\% \\ 
        $\beta_{15}$ & 0.007 & 0.009 & 0.008 & 95.5\%  & 0.006 & 0.009 & 0.009 & 94.5\% \\ 
        $\beta_{16}$ & 0.009 & 0.006 & 0.005 & 92.0\%  & 0.007 & 0.008 & 0.009 & 95.0\% \\ 
        $\beta_{17}$ & 0.008 & 0.006 & 0.006 & 94.5\% & 0.008 & 0.008 & 0.008 & 95.5\% \\ 
        $\beta_{18}$  & 0.008 & 0.009 & 0.008 & 95.0\%  & 0.009 & 0.009 & 0.009 & 94.5\% \\ 
        $\beta_{19}$ & 0.008 & 0.008 & 0.008 & 95.0\% & 0.007 & 0.008 & 0.008 & 95.0\% \\ 
        $\beta_{20}$ & 0.008 & 0.007 & 0.009 & 93.5\% & 0.008 & 0.009 & 0.009 & 94.5\% \\ 
        $\eta$  & 0.062 & 0.052 & 0.052 & 95.5\% & 0.066 & 0.055 & 0.055 & 95.0\% \\ \hline
    \end{tabular}}
    \label{sim3_2}
\end{table}

\newpage

\begin{table}[!ht]
    \centering
    \captionsetup{font={small,bf,stretch=1.25},justification=raggedright}
    \caption{Specification of hyperparameters in Analysis of HER2-positive Breast Cancer Spatial
Transcriptomic Data}
    \begin{tabular}{c|c}
    \hline
        \textbf{Parameters} & \textbf{Values} \\ \hline
        $b_1$ & 5\\ 
        $b_2$ & 50\\ 
        $b_3$ & 5\\ 
        $b_4$ & 50\\
        $b_5$ & 0\\ 
        $b_6$ & 10\\ 
        $c_1$ & 8\\ 
        $v_0$ & 0.000001\\ \hline
    \end{tabular}
    \label{real.data.hyperpara}
\end{table}


\bibliographystyle{apa-good}
\clearpage
\phantomsection  
\renewcommand*{\bibname}{References}

\addcontentsline{toc}{chapter}{\textbf{References}}

\bibliography{reference}

\nocite{*}



        



\end{document}